\documentclass[smallextended]{svjour3}       
\smartqed  
\journalname{Empirical Software Engineering}

\usepackage{listings}
\usepackage{graphicx}
\usepackage{hyperref}
\usepackage{subcaption}
\usepackage{supertabular}
\usepackage{paralist}
\usepackage{amsfonts}
\usepackage{amsmath} 
\usepackage{tcolorbox}

\usepackage{verbments}
\usepackage{verbatim}
\usepackage{booktabs} 
\usepackage{ifthen}
\usepackage{color}
\usepackage{xspace}
\usepackage{tikz}
\usetikzlibrary{fit,arrows,calc,positioning}
\usetikzlibrary{positioning}
\newboolean{showcomments}
\setboolean{showcomments}{true} 
\ifthenelse{\boolean{showcomments}}
  {\newcommand{\nb}[3]{
  {\color{#2}\small\fbox{\bfseries\sffamily\scriptsize#1}}
  {\color{#2}\sffamily\small$\triangleright~$\textit{\small #3}$~\triangleleft$\GenericWarning{}{LaTeX Warning: #1: #3}}
  }
  \newcommand{\todo}[1]{{\color{red}{TODO: #1}}\GenericWarning{}{LaTeX Warning: TODO: #1}}
  }
  {\newcommand{\nb}[3]{}
  \newcommand{\todo}[1]{}
  }

\newcommand{\pef}{Observation-based\xspace}
\usepackage{color}
\usepackage{colortbl}
\definecolor{mycolor3}{cmyk}{0, 0.708, 0.49, 0.112}
\definecolor{aliceblue}{rgb}{0.64, 0.76, 0.68}

\newtcolorbox{mybox}[3][]
{
  colframe = #2!25,
  colback  = #2!10,
  coltitle = #2!20!black,  
  #1,
}

\begin{document}

\title{E-APR: Mapping the Effectiveness of Automated Program Repair Techniques}

\author{Aldeida Aleti         \and
       Matias Martinez 
}


\institute{A. Aleti \at
              Faculty of Information Technology, Monash University, Australia. \\
              \email{aldeida.aleti@monash.edu}           
           \and
           M. Martinez \at
              Universit\'e Polytechnique Hauts-de-France, France. \\
              \email{matias.martinez@uphf.fr}
}

\maketitle
\begin{abstract}
Automated Program Repair (APR) is a fast growing area with numerous new techniques being developed to tackle one of the most challenging software engineering problems. APR techniques have shown promising results, giving us hope that one day it will be possible for software to repair itself. In this paper, we focus on the problem of objective performance evaluation of APR techniques. We introduce a new approach, Explaining Automated Program Repair (E-APR), which identifies features of buggy programs that explain why a particular instance is difficult  for an APR technique. E-APR is used to examine the diversity and quality of the buggy programs used by most researchers, and analyse the strengths and weaknesses of existing APR techniques. E-APR visualises an instance space of buggy programs, with each buggy program represented as a point in the space. The instance space is constructed to reveal areas of hard and easy buggy programs, and enables the strengths and weaknesses of APR techniques to be identified.

\end{abstract}

\section{Introduction}\label{sec:introduction}

Software can not be seen or touched, but it has a physical existence. With software embedded into many devices today, software failures have caused not only inconveniences but also tragedies, such as the deaths of patients due to massive overdose caused by an avoidable error in a radiation therapy machine~\cite{kaner2008lessons}. 
A more recent case is Google’s self-driving cars (controlled by software), which experienced 272 failures in less than a year. These failures would have resulted in at least 13 crushes killing their human drivers if they had not intervened~\cite{selfdrivingcar}. Software failures are also the cause of massive economical losses, costing the global economy \$41 billion annually~\cite{failurecost}. 
Repairing software faults, however, is becoming an extremely difficult and expensive task -- constituting up to 90\% of the software expenses \cite{LeGoues2013current} -- due to the increasing complexity and size of software systems. A modern car, for example, has 100 million lines of code, and this number is expected to increase to 200-300 millions in the near future~\cite{linesofcode}. Hence the critical task of software repair must be automated. 

Automated Program Repair (APR) has been identified as the grand challenge in software engineering research~\cite{FIFVerify}. Many APR methods have shown promising results in fixing bugs with minimal, or even no human intervention~\cite{LeGoues2012GenProg,LeGoues2012Study,MartinezM15,Xuan16MDCLDLM}. Despite many studies introducing various APR techniques (APRTs), much remains to be learned, however, about what makes a particular technique work well (or not) for a specific software system~\cite{anand:2013}. The effectiveness of APRTs is likely to be problem dependent, which calls for an analysis of the software characteristics that impact their effectiveness in order to help practitioners select the most appropriate technique for their software system.

In addition, results claiming the superior performance of an APRT over other techniques on a selected set of software systems may not generalise to untested systems. It is likely that there are software systems where an APRT excels because it is exploiting some particular characteristics of the buggy program. Thus, an understanding of conditions under which an APRT can be expected to succeed or fail is essential, however, this is rarely included in published studies. The aim of this paper is to address the issue of objective assessment of APRTs, and we achieve this by answering the following research questions:

\begin{itemize}
\item[ ]\textbf{RQ1} What impacts the effectiveness of APRTs? - Research introducing new APRTs or experimental studies investigating the performance of different techniques usually is based on a carefully selected set of buggy programs. These works offer little insight into the characteristics of the buggy programs and how they are correlated with the effectiveness of APRTs. The overwhelming majority of published work in APR only describes the benefits of the newly introduced technique, while just a few mention the limitations or present negative results. 

Certain limitations of APRTs have previously been discussed in the literature, such as the issue with patch overfitting (a patch generated by a tool that, while being valid according to the correctness oracle, they are still incorrect and potentially introduce new bugs that can not be captured by the correctness oracle). On the other hand, negative results in terms of why certain techniques can not repair certain bugs have not been investigated in the literature so far. In this paper, we aim to find out if particular features of a buggy program correlate with the effectiveness of APRTs, thus providing insights on why some techniques might be more or less suited to certain software and bug instances. We achieve these kind of insights by proposing a new method for analysing the effectiveness of APRTs.

\item[ ] \textbf{RQ2} Are APR datasets significantly different? - Most research in APR uses well-known datasets, such Defects4J, which can result in the techniques to be tailored towards solving particular problems, and as a result not generalise well for other problems. In this paper, we aim to show how different these datasets are in terms of the features that have an impact on the effectiveness of existing APRTs. This allows us to understand if existing benchmarks are sufficiently diverse for stress-testing the effectiveness of APR techniques.
 
\item[ ]\textbf{RQ3} How can we select the most suitable APR technique? The final aim of this research is to investigate the effectiveness of Machine Learning techniques for APRT selection. We investigate different multi-label classification techniques and report their effectiveness in terms of recall, precision and f1-score.

\end{itemize}

To answer these research questions, we introduce a new approach which characterises both strengths and weaknesses of existing APR techniques. E-APR is inspired from earlier work on instance space analysis in the area of machine learning~\cite{munoz2018instance} and search based software testing (SBST)~\cite{OlivieraAleti,oliveira2019footprints}. These approaches are concerned with the problem of objective performance evaluation of different algorithms used in machine learning~\cite{munoz2018instance} and SBST~\cite{OlivieraAleti,oliveira2019footprints}, and the impact of the choice of problem instances. The methodology used in these studies extend the Rice's framework~\cite{rice1976algorithm} with the aim of gaining insights into why some algorithms might be more or less suited to certain problem instances.

E-APR extends the  methodology from Oliveira et al.~\cite{OlivieraAleti,oliveira2019footprints} and Munoz et al.~\cite{munoz2018instance} to the automated program repair problem. E-APR allows for a more objective assessment of existing APR techniques, and helps in understanding why certain APR techniques cannot generate plausible patches for certain bugs. We apply our framework on a large study of 2,141 bugs from 130 projects, and 23,551 repair attempts. E-APR uses software and bug features to characterise the buggy program instances, and learns which features have an impact on the effectiveness of APRTs. For human programmers, software repair is challenging because fixing bugs is a difficult task. While there are bugs that can be trivially fixed, many of us can remember a bug that took hours, if not days and weeks to be understood and fixed~\cite{eisenstadt1997my}. The approach we devise gives insights into how an APR technique can be selected to automatically fix bugs.

\section{The E-APR Framework}\label{sec:conceptualframework}

The E-APR framework has two main goals:
\begin{itemize}
    \item to help designers of APRTs gain insight into why some techniques might be more or less suited to repair certain buggy programs, thus devising new and better techniques that address any challenging areas, and
    \item to help software developers select the most effective APRT for their software system.
\end{itemize}

E-APR provides a way for objective assessment of the overall effectiveness of an APR technique. It is based on previous work on instance space analysis and algorithm selection in the area of Search-Based Software Testing~\cite{oliveira2019footprints,OlivieraAleti}, machine learning~\cite{munoz2018instance}, and optimisation~\cite{smith2012measuring}. The concept of instance space analysis was first introduced by Smith-Miles in her seminal work looking at the strengths and weaknesses of optimisation problems, and forms the foundation of the E-APR approach. Understanding the effectiveness of an APR technique is critical for selecting the most suitable technique for a particular buggy program, thus avoiding trial and error application of APR techniques. 

 
An overview of the E-APR framework is presented in Figure~\ref{fig:method}. E-APR starts with a set of buggy programs $p\in P$ and a portfolio of APRTs $t\in T$. The performance of APRTs is measured for each buggy program as $y(t,p)$, which indicates whether a plausible patch has been found for that program. The first step of E-APR is to identify the significant features of buggy programs ($f(p)\in F$) that have an impact on how easy or hard they are for a particular APRT. Next, E-APR constructs the APRT footprints $(g(f(P)))\in R^2$ which indicate the area of strength for each APRT. Finally, E-APR applies machine learning techniques on the most significant features to learn a model that can be used for APRT selection for future application.

\tikzstyle{block} = [rectangle, draw, fill=mycolor3!15,text width=9em,  text centered, minimum height=4em, rounded corners=4pt]

\tikzstyle{block1} = [rectangle, draw, fill=white,text width=9em,  text centered, minimum height=4em, rounded corners=4pt]

\tikzstyle{big} = [rectangle, draw, inner sep=0.5cm]

\tikzstyle{line} = [draw, -latex']
\tikzstyle{node}=[rectangle,draw=grey,text width=3cm,inner sep=0.2cm,text centered]

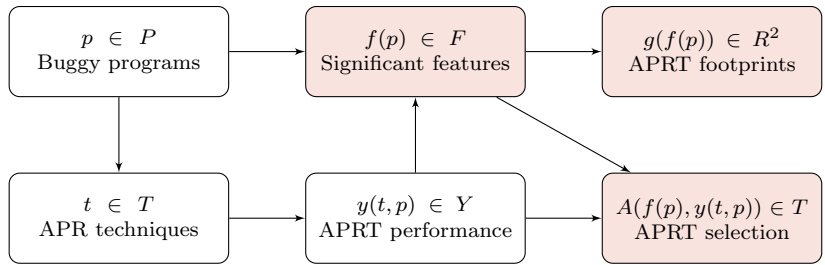
\begin{figure}
 \centering
 \begin{tikzpicture}
\node [block1](1) {\textbf{$p\in P$} \\ Buggy programs};
\node [block, right=of 1] (2){$f(p) \in F$ \\ Significant features};
\node [block, right=of 2] (3) {\textbf{$g(f(p)) \in R^2$ \\} APRT footprints};

\node [block1, below=1cm of 1] (4) {$t \in T$  \\ APR techniques};
\node [block1, right=of 4] (5) {\textbf{$y(t, p) \in Y$ \\} APRT performance};
\node [block, right=of 5] (6) {$A(f(p), y(t,p)) \in T$ \\ APRT selection};

\path [line] (1)--(2);
\path [line] (2) -- (3);
\path [line] (1)--(4);
\path [line] (4)--(5);
\path [line] (5) -- (2); 
\path [line] (2) -- (6); 
\path [line] (5) -- (6); 
\end{tikzpicture}
\caption{\normalsize An overview of E-APR.\label{fig:method}}
 \end{figure}
 
\subsection{Buggy Programs}\label{sec:bug_benchmark}

Buggy Programs, defined in Figure~\ref{fig:method} as $p\in P$ are software instances used by researchers to evaluate automated program repair techniques. Most of the APRTs for Java use Defects4J \cite{Just2014defects4j}.
Durieux et al.~\cite{Durieux:2019:RepairThemAll} is one of the few that uses 5 peer-reviewed Java bug benchmarks: Bears \cite{Madeiral2019Bears}, Bugs.jar \cite{Saha2018BugsDotjar}, IntroClassJava and QuixBugs \cite{lin2017QuixBugs} and Defects4J~\cite{Just2014defects4j}. Our analysis is based on the experimental data generated by Durieux et al.~\cite{Durieux2016IntroClassJava}, which is available at \url{github.com/program-repair/RepairThemAll_experiment}.

In total, we consider 2,141 bugs from 130 projects, and 23,551 repair attempts. A repair attempt is the execution of an APRT on a buggy program. The execution of all repair attempts on the 5 benchmarks by the 11 APRTs took 314 days \cite{Durieux:2019:RepairThemAll}. The patches considered in this study are \emph{plausible patches}. These patches produce: a) the failing test cases (that exposed the bug) pass,  and b) the remaining test cases continue to pass. Those patches are also known as \emph{plausible} patches \cite{Qi2015Kali}.
Previous work have shown that a test-suite adequate patch can produce passing all tests but they are yet incorrect. Those are \emph{overfitting} patches \cite{Smith2015} and can arise due to the weakness of the test-suite used for synthesising the patches. Overfitting detection is not yet mature (i.e., not capable of detecting all overfitting patches) and thus adopting such techniques could introduce some bias in this work, hence we consider \emph{all} patches generated by the repair tools executed by RepairThemAll.
This means that we did not filter out the outputs generated by APRTs.

The source of the bugs in the bug benchmark are diverse: Defects4J and Bugs.jar contains real bugs extracted from software repositories, 
Bears contains real bugs collected from breaking builds on Travis platforms, IntroClassJava contains buggy subjects from students, and QuixBugs contains buggy implementation of well-known algorithms (such as merge-sort). 

\subsection{APR Techniques}
\label{sec:all_approaches}
APR techniques are defined in Figure~\ref{fig:method} as $t\in T$. In this paper we focus on one family of repair approaches: \emph{test-suite based repair approaches} \cite{LeGoues2012GenProg}.
Approaches from this family aim at repairing bugs exposed by at least one failing test case. The main idea of these approaches is to use failed test cases to localise potential faults and then apply mutations to the source code until the program satisfies all unit test cases. The mutations that are applied to the program code can range from small changes like modification, addition or removal of a single code line~\cite{LeGoues2012GenProg} to complex edit operations~\cite{MartinezM15,Kim13PAR}, which are mined from software repositories and used to fix a fault in a different context. 


In this paper we employ 11 repair tools for Java programs similar to the study by Durieux et al. \cite{Durieux:2019:RepairThemAll}. These tools can be classified into \emph{semantics-based repair tools} (Nopol~\cite{Xuan16MDCLDLM} and DynaMoth\cite{Durieux2016DynaMoth}), \emph{a metaprogramming-based tool} (NPEFix \cite{Durieux2017NPEFix}), and \emph{generate-and-validate} (ARJA \cite{Yuan2018ARJA}, Cardumen \cite{Martinez2018Cardumen}, jGenProg \cite{Martinez2016Astor}, GenProg-A \cite{Yuan2018ARJA}, jKali \cite{Martinez2017experiment},
Kali-A \cite{Yuan2018ARJA}, jMutRepair \cite{Martinez2016Astor}, and RSRepair-A \cite{Yuan2018ARJA}).

\textbf{jGenProg and GenProg-A} are two Java implementations of GenProg~\cite{LeGoues2012GenProg}. Both techniques use a generate-and-validate method to produce patches using a genetic programming approach. The search space consists of patches that are formed through combinations of removing code, and inserting and replacing code from elsewhere in the program under repair~\cite{Martinez2016Astor}. 

\textbf{Cardumen}~\cite{Martinez2018Cardumen} synthesises patches using the existing code as a basis, by taking code elements from elsewhere in the program and replacing the variables. Each potential patch is filtered based on location and type compatibility, and the remaining patches are prioritised based on how frequently the selected variables occur together. 

\textbf{jKali and Kali-A}~\cite{Qi2015Kali} are different implementations of Kali in Java. They attempt to come up with candidate patches by removing or skipping statements. Neither jKali nor Kali-A is a 'repair' program, instead, they are more useful in identifying weak test suites and under-specified bugs~\cite{martinez2017automatic}. Since Kali simply removes or skips code, if a patch is found, it is a strong indication that the functionally of the removed code is not specified in the test-suite. In addition, if Kali finds a test-suite adequate patch, so can jGenProg or Nopol~\cite{martinez2017automatic}, the patches found by Kali, however, rarely work beyond the given test-suite.

\textbf{jMutRepair}~\cite{Martinez2016Astor} performs an exhaustive search of the code and applies the following three types mutation operators on suspicious if conditions. The relational mutation operator with the following values (==,!=,$\leq$,$\geq$,$\textless$,$\textgreater$), the logical mutation operator (AND, OR), and the \textit{Unary} mutation operator which applies negation and positivation.

\textbf{Nopol}~\cite{Xuan16MDCLDLM} focuses on repairing IF conditions, which are amongst the most error-prone elements of Java programs, and many one-change commits simply update an IF condition. Nopol has three main steps. First, it locates a fix location for a potential patch using ``angelic fix localisation''. This process also involved finding ``angelic values'', which are assigned values that can be used at the fix location to make all failing tests pass. Next, Nopol collects runtime data from a test execution, including a snapshot of the program state at candidate fix locations. Then, Nopol translates the angelic values and available variables at the fix location into a Satisfiability Modulo Theorem problem, and attempts to find a solution, which is then translated into a patch. 

\textbf{RSRepair-A}~\cite{Qi2014RSRepair} is a Java implementation of the RSRepair program repair tool written for C programs. RSRepair uses a generate-and-validate technique to prepare patches. It takes inspiration from the GenProg tool, however, instead of using genetic programming as its search method, RSRepair uses random search. 

\textbf{ARJA}~\cite{Yuan2018ARJA} uses Genetic Programming to modify and mutate suspicious statements in a program by performing three actions: i) deleting the suspicious statement, ii) replacing the suspicious statement, or iii) inserting extra statements before or after the suspicious statement. ARJA reduces the scope of the search and computation time to speed up the fitness process by applying rules that exclude statements that are not related to the problem~\cite{Yuan2018ARJA}.

\textbf{NPEFix}~\cite{Durieux2017NPEFix} repairs null pointer exceptions at runtime by using two strategies. The first strategy assigns an alternative value (which can be a valid value that is stored in another variable or a random value) for a null dereference.  The second strategy skips the execution of the null dereference, by either skipping a single statement or skipping the complete method. All strategies are applicable for any arbitrary objects, including instances of library classes, and instances of domain classes.

In summary, the APR techniques discussed in this section can be broadly categorised based on their high-level repair strategy. For example, jGenProg~\cite{Martinez2016Astor}, ARJA~\cite{Yuan2018ARJA} and RSRepair-A~\cite{Qi2014RSRepair} use or build upon genetic programming. Other techniques take more unique approaches and are designed to target specific bugs, like NPEFix~\cite{Durieux2017NPEFix} targeting null pointer exceptions. Other repair tools can only function if code is structured in a certain way, like Nopol~\cite{Xuan16MDCLDLM}, which only works when IF conditions are present, and will only find a valid patch if the patch involves changing IF conditions. These observations further support our hypothesise that the performance of each technique will likely be affected by the features of the code. Different repair strategies may favour different code features, and that different bug targeting will definitely perform badly on code with the wrong type of bug.

\subsection{APRT Performance Measures}

An APRT Performance Measures $y(t,p) \in Y$ takes as input the patches generated by an APRT $t\in T$ for a particular buggy program $p\in P$. There exist various measures of APRT performance focusing on the quality of the patches produced. In this work we consider \emph{test-suite adequate} patches~\cite{LeGoues2012GenProg}.
We acknowledge that a portion of the patches may be overfitting, i.e., according to the test suite, the buggy program may appear to have been fixed by the patch, however, new errors may have been introduced. The problem of filtering correct patches (e.g., \cite{Yu2019Alleviating,Xiong2018Identifiying,issta17-difftgen,Le2019Reliability}) is currently being addressed by many researchers, who are looking at ways for automating or semi-automating this process, since manually inspecting all generated patches by automated program repair techniques is not practical. The APRT performance measures in E-APR ($y(t,p)\in Y$) can easily be extended to new measures, such as patch correctness.

\subsection{Significant Features} 
\label{sec:framework:featlearning}

A critical step of E-APR is identifying features of buggy program instances $f(p) \in F$ that have an impact on the effectiveness of APR techniques. Features are problem dependent and must be chosen such that the varying complexities of the buggy program instances are exposed, any known structural properties of the software systems are captured, and any known advantages and limitations of the different APRTs are related to features. 

For the purpose of this work, an APR technique is effective if it can generate a plausible patch for a buggy software system. While much is known and reported on features that correlate with software quality, we must consider that there may be other unknown features that have an impact on the effectiveness of APR techniques. In addition, it is possible that not all known features are useful for our goal of separating the hard and easy software instances. The candidate set of features may contain redundancy, with features measuring aspects of a buggy program that are either similar or not relevant to expose the hardness of the APR task itself. Thus, a small set of relevant features must be selected.


Learning significant features has two steps: first we define how to measure the quality of a particular set of features, and second, we apply a Genetic Algorithm to select the set that maximises this measure. A subset of features is considered of high quality if they result in an instance space -- as defined by the 2-dimensional projection of the subset of features -- with buggy programs that show similar performance of APRTs closer to each other. The best subset of features is the one that can best discriminate between easy and hard buggy program instances for APR techniques.

E-APR aims at identifying features that are able to create a clear separation of the buggy program instances, such that we can clearly see the different clusters of buggy programs where each APRT is effective. We employ principal component analysis (PCA)~\cite{jolliffe2011principal} to locate significant features. PCA learns a linear combinations of the buggy program features. The first PC is the linear combination of the variables which explain the maximum amount of variance in the dataset. Each subsequent PC is orthogonal to all previously calculated PCs and captures a maximum variance under these conditions. In our work, the subset of variables that have large coefficients and therefore contribute significantly to the variance of each PC, are identified as the significant features which are selected to explain bugs.

Given $|F|$ software features, we can have at most $|F|$ components which are estimated in decreasing order of the variance (measured through the eigenvalue of each PC) they explain in the dataset. We analyse for each PC the features that are found significant. This shows which dimensions are the main drivers of APR technique effectiveness and help explain why this is the case. In PCA, usually only the first few components are regarded as important. We retain the first 2 components, which makes visualising the footprints of the algorithms much easier.

E-APR uses a genetic algorithm~\cite{aleti2014choosing} to search the space of possible subsets of $k$ features, with the classification accuracy on an out-of-sample test set used as the fitness function to guide the search for the optimal subset. The instance space is generated in iterations, until an optimal subset of features is found~\cite{munoz2018instance}. The genetic algorithm performs the following steps to select the features and generate the instance space:
\begin{enumerate}
    \item a set of buggy program features is selected;
    \item an instance space is generated using the selected features and PCA to reduce the dimensionality; 
    \item the fitness of the set of features is evaluated ;
\item if the features are not adequate, go back to step 1.
\end{enumerate}

Once the best set of features features is identified, E-APR creates a 2-D instance space that helps inspect the relationships between problem instances, their features and objectively assess APRT performance. 2D visualisation has been found to be effective in visualising footprints~\cite{oliveira2019footprints,OlivieraAleti,munoz2018instance}, hence we follow a similar approach as previous work. Similar approaches have been proposed in the literature for feature subset selection for machine learning~\cite{bengio2003extensions}, optimisation~\cite{smith2014towards}, and search-based software testing tasks~\cite{OlivieraAleti}. Certainly, other feature selection methods proposed in the literature~\cite{guyon2003introduction} would also be suitable for the task at hand.

\subsection{APRT Footprints}
\label{sec:framework:visualisation}
The idea of algorithm footprints was first introduced by Smith-Miles and Tan~\cite{smith2012measuring} and aims to determine the relative performance of different algorithms across various classes of instances. In the original paper~\cite{smith2012measuring}, the authors focused on optimisation problems. Rather than reporting algorithm performance averaged across a chosen set of benchmark instances, the authors develop metrics for an algorithm’s performance generalised across a diverse set of instances. E-APR extends these ideas to Automated Program Repair techniques and aims to measure APRT footprint, which gives an indication of the area of strength of these algorithms.

Once the significant features have been identified, they are used to analyse and visualise the footprints of the APR techniques. In order to facilitate the visualisation of the footprints, similar to previous work~\cite{smith2012measuring,oliveira2019footprints,OlivieraAleti}, we utilise the 2-D instance space created using PCA as a dimensionality reduction technique, and project the instances to two dimensions, while making sure that we retain as much information as possible. PCA rotates the data to a new coordinate system $\mathbf{R}^k$, with axes defined by linear combinations of the selected $F^*$ features, where $k=|F^*|$. The $k$ new axes are the eigenvectors of the $k\times k$ covariance matrix. 

We retain the two principal eigenvectors which correspond to the two largest eigenvalues of the covariance matrix. The instance space is then projected on this two-dimensional space. We use the variance explained in the data by the two principal components as a measure of the loss in information due to dimensionality reduction. Following a similar approach to previous work on dimensionality reduction~\cite{smith2014towards}, we accept the new two dimensional instance space as adequate if most of the variance in the data is explained by the two principal axes. The two principal components $z_1$ and $z_2$ are then used to visualise the footprints of the APR technique (APRT).

If our goal was only to make performance predictions on the best APR tool for repairing a particular software system, we could use machine learning algorithms to identify the relationship between software features and APR performance. Machine learning on its own does not allow for explanations as to why a particular APRT works well. Our goal in this paper is much broader than only making prediction, as we aim to visualise the footprints of the different APR approaches and provide insights into the workings of these methods.

Next, we calculate the relative size of APRT footprints by estimating the area of the hull covering the software instances where the technique is expected to perform well. This is a metric of the relative goodness of the APRT across the software instance space. Formally, given the convex hull $H(S)$ of an area defined by points $S=\{(x_i,y_i)\}, \forall i=1,...,n$, the area $A(H(S))$ is given by
\begin{equation}\label{eqFootprintsize}
    A(H(S))=\frac{1}{2} \sum_{j=1}^k(x_jy_{j+1}-y_jx_{j+1})+(x_ky_1-y_kx_1),
\end{equation}
where the subset $\{(x_j,y_j), \forall j=1,...,k\}, k\leq n$ defines the extreme points of $H(S)$. Using Equation~\ref{eqFootprintsize}, we compare the relative size of the footprint of each APRT to determine which APRT has the largest footprint and explore the degree of overlap of the footprints. 

\subsection{APRT Selection}
\label{sec:framework:selection}

In the final step, E-APR predicts, based on the most significant software features, the most effective APR technique for repairing particular buggy programs. E-APR uses the most significant features as an input to machine learning algorithms to learn the relationship between the instance features and APR method performance. For this purpose, we can use a variety of machine learning algorithms, such as decision trees, or support vector machines for binary labels (effective/ineffective), or statistical prediction methods, such as regression algorithms or neural networks for continuous labels (e.g., time complexity of the approach). 

In this work, we investigate four machine learning approaches for multi-label classification~\cite{madjarov2012extensive}. These methods are support vector machine (SVM)~\cite{Boser1992SVM}, a random forest classifier (RFC)~\cite{prabhu2014fastxml}, a decision tree (DT)~\cite{quinlan1996learning} and a multi-layer perceptron (MLP)~\cite{ruck1990multilayer}. 
We now briefly describe those techniques.

\textbf{Support Vector Machine (SVM)} 
is a supervised learning model with associated learning algorithms that analyse data for classification and regression analysis.
For classification, SVM aims at finding a hyper-plane in the feature space, which separates the training data into two classes while maximising the margin (in the feature space) between this hyper-plane and the two classes \cite{Vapnik1995}.

\textbf{Decision Tree (DT)} uses observations about an item (represented in the branches) to learn an item's target value (represented in the leaves). Classification trees are those trees where the target variable can take a discrete set of value, leaves represent class labels and branches represent conjunctions of features that lead to those class labels.

\textbf{Random Forest Classifier (RFC)}
is an ensemble learning method for classification that operates by constructing a multitude of decision trees at training time and outputting the class that is the mode of the classes.

\textbf{Multi-Layer Perceptron (MLP)} consists of a feed-forward artificial neural network which is a system of interconnected neurons representing a nonlinear mapping between an input vector and an output vector. MLP is used for classification by assigning output nodes to represent each class. MLPs are typically trained using a supervised learning technique called back-propagation.

At the end of this process, E-APR produces a model that can be used for algorithms selection in automated program repair. This model can be retrained and extended with more APR tools and features.

\section{Experimental Design}

We implement the E-APR framework described in Section \ref{sec:conceptualframework}, and conduct a set of experiments and analysis to answer the research questions stated in Section~\ref{sec:introduction}. In this section, we describe:
the automated program repair techniques, the benchmark of buggy programs, and the set of software features.

\subsection{Buggy Program Features}
\label{sec:method:features}

Features are problem dependent and must be chosen so that the varying complexities of the problem instances are exposed, any known structural properties of the buggy programs are captured, and any known advantages and limitations of the different program repair techniques are related to features. The most common measures and metrics used to characterise features of a software system are extracted from code.

Among others, we use object-oriented code metrics based on measurement theory and expertise of experienced software developers \cite{chidamber:1994}. These metrics are also mapped to the Quality Model for Object-Oriented Design~\cite{el2004object}, which is a comprehensive model that establishes a clearly defined and empirically validated model to assess object-oriented design quality attributes such as understandability and reusability, and relates them through mathematical formulas with structural object-oriented design properties such as encapsulation and coupling. 
The set of code metrics, {presented in Table \ref{tab:featuresObject}}, includes simple metrics, which count the number of methods or lines of code, to more complex metrics that measure the interaction between methods and the depth of inheritance tree. 
As in this paper we focus on Java APR, we also include Java-Specific method features, which are presented in Table \ref{tab:featuresJava}.

\begin{table}[!ht]
    \caption{Object-oriented features.}
    \label{tab:featuresObject}
\renewcommand{\arraystretch}{1.1}
    \begin{tabular}{l|p{10cm}}
    \hline
 WMC&Weighted methods per class is a measure of complexity in a class~\cite{Chidamber94}. \\
 DIT& Depth of inheritance is the depth of inheritance of the class (i.e. number of ancestors in direct lineage)~\cite{Chidamber94}.\\
 NOC& Number of children is an indication of the scope of properties. It counts the sub-classes that inherit the methods of the parent class~\cite{Chidamber94}.\\
 CBO& Coupling between object classes  is a count of the number of other classes to which the current class is coupled~\cite{Chidamber94}.\\
 RFC& Response for a class measures the interaction of the class' methods with other methods~\cite{Chidamber94}.\\
 LCOM& Lack of cohesion in methods. This metric counts the sets of methods in a class that are not related through the sharing of some of the class's fields~\cite{Chidamber94}.\\
 CA& Afferent coupling is a measure of how many other classes use the specific class.\\
 CE& Efferent couplings. This is a measure of how many other classes are called within the given class.\\
 LCOM3& Lack of cohesion in methods. This metric is defined as the number of connected components in the call graph.\\
 NPM& Number of public methods for a class\\
 LOC& Lines of code. As the name indicates, this measure counts the lines of code in a class. We take the average lines of code per class in a buggy program.\\
 DAM& Data access metric. This metric is the ratio of the number of private and protected attributes to the total number of attributes declared in the class.\\
 MOA& Measure of aggregation. This is the percentage of data declaration in the system whose types are of user defined classes (i.e., data types other than system defined classes such as integers, real numbers etc).\\
 MFA& Measure of functional abstraction is the ratio of the number of methods inherited by a class to the total number of methods accessible by members in the class.\\
 CAM& Cohesion among methods of class computes the relatedness among methods of a class based upon the parameter list of the methods.\\
 IC& Inheritance coupling calculates the number of parent classes to which a given class is coupled. \\
 CBM& Coupling between methods measures the total number of new/redefined methods to which all the inherited methods are coupled.\\
 AMC& Average method complexity measures the average method size (the number of java binary codes in the method) for each class.\\
 \hline
 \end{tabular}
 \end{table}
 
 \begin{table}[!ht]
    \caption{Java specific method features.}
    \label{tab:featuresJava}
\renewcommand{\arraystretch}{1.1}
    \begin{tabular}{l|p{10cm}}
    \hline
	AC& Abstract methods count is the number of abstract methods in a class.\\
	ASMC& Abstract static methods count is the number of static methods in a class.\\
	 DAMC& Default abstract methods count.\\
     DASMC& Default abstract static methods count.\\
	DMC& Default methods.\\
	 DSM& Default static methods count.\\
	 GMC& General methods count\\
	 GSMC& General static methods count\\
	 MC& Methods count.\\
	 PriAMC& Private abstract methods count.\\
	 PriASMC& Private abstract static methods count.\\
	 PMC& Private methods count.\\
	 PSMC& Private static methods count.\\
	 ProAMC& Protected abstract methods count.\\
	 ProASMC& Protected abstract static methods count.\\
	 ProMC& Protected methods count.\\
	 ProSMC& Protected static methods count.\\
	 PubAMC& Public abstract methods count. \\
	 PubASMC& Public abstract static methods count.\\
	 PubMC& Public methods count.\\
	 PubSMC& Public static methods count.\\
	 SMC& Static methods count\\\hline
    \multicolumn{2}{c}{\textbf{\pef{} Features (see complete list at \cite{Yu2019XCRF})}}\\\hline
	Usage& Related to usage of e.g. variables and invocations\\
	
	\hline
	Syntax& Related to syntax of e.g. variable's identifiers \\
	
	\hline
	Types& Related to types of e.g. variables, and parameters.\\
    \end{tabular}

\end{table}

In addition to code features widely used by software practitioners and researchers, we also consider a set of \pef features \cite{Yu2019XCRF}.
Those features were manually crafted for targeting different open challenges from  automated program repair's field such as prediction of source code transformations on buggy code \cite{Yu2019XCRF} and detection of incorrect patches \cite{Ye2019ODS}.

\pef features capture different characteristics of a buggy program.
Initially, Defects4J \cite{Just2014defects4j} was considered as a starting point, which is a dataset of real Java bugs and the corresponding human-written patches, widely used in evaluations of automated program repair tools \cite{Durieux:2019:RepairThemAll}.
We recorded the following information for each code element in the buggy code affected by the patch and in the patched code:
\begin{inparaenum}[\it a)]
\item the characteristics of the elements (e.g., the type of a variable is primitive), and
\item the relation of such elements with respect to the rest of the buggy and patched file, respectively. 
\end{inparaenum}
Finally, the designers defined a set of features from those observations.
For example, from Listing \ref{listingDiff1}, it was observed that the buggy statement references to a variable ($p1$) which has compatible type and similar name to another variable ($p2$) in scope. 
From this observation, they created a feature named ``HVSN'' (Has Variable with Similar Name).


\begin{lstlisting}[language=java, label=listingDiff1, caption=Human-written for bug Chart-11 from Defects4J.]
 @@ -272,7 +272,7 @@ public static boolean equal(
        GeneralPath p1, GeneralPath p2)    
             
        PathIterator iterator1 = p1.getPathIterator(null);            
-       PathIterator iterator2 = p1.getPathIterator(null);            
+       PathIterator iterator2 = p2.getPathIterator(null);

\end{lstlisting}

\pef features were included in the set of buggy program features because they allow us to capture the characteristics of code elements related to bug that: 
\begin{inparaenum}[\it a)]
\item can be repaired by a tool, and
\item cannot be repaired by any tool.
\end{inparaenum}
Thus, our approach could predict whether a buggy program can be repaired (or not) by a particular repair tool, and to determine which is the most adequate repair tool to face the bug.
A simple example to illustrate the intention behind the adoption of such features: the buggy version of bug Chart-11 from Listing \ref{listingDiff1} has the feature $HVSN$ with a $true$ value and it is successfully repaired by Cardumen \cite{Martinez2018Cardumen} but neither by for instance Arja nor GenProg \cite{Durieux:2019:RepairThemAll}.
Thus, our intuition is that other bugs having that feature could be repaired by Cardumen.

\pef{} features are grouped into three categories: 
\begin{inparaenum}[\it 1)]
\item features related to the \emph{Usage} of code elements, for example, the feature OUIA indicates if a statement references a local variable that has not been referenced in other statements before it, 
\item features related to the \emph{Syntax} of code elements, for example, 
the feature HVSN (Has Variable with Similar Name) indicates whether, given a statement that references a variable, there exist other variables in the same scope that have a similar identifier name with that variable; 
\item features related to the \emph{Types} of code elements, for example, the feature VTSV indicates whether, given a statement that references a variable, there exist other variables in the same scope that are type compatible with that variable.
\end{inparaenum}
In total, we have 146 \pef{} features. 
The complete list is available in our appendix~\cite{appendix}.
These features can be computed using the open-source tool Coming \cite{Martinez2019Coming}, which is available online at \url{https://github.com/SpoonLabs/coming}.

\subsection{E-APR Input Data}

For each buggy program, we first create a vector where each dimension corresponds to a particular feature. We add to that vector an additional dimension per each APRT considered in this experiment: its value is `1' if the corresponding APRT produced a plausible patch and a `0' otherwise. Table \ref{tab:dataset} shows an example of the features extracted from 4 buggy programs. Each row has the values of the features extracted for a program, and it is a vector of features. 
From the second to the fifth column, it shows the values corresponding to 4 object-oriented features (wmc, dit, npc and cbo). 
The last two columns indicate whether the buggy program could be repaired by two approaches (Kali and Arja).

\begin{table}[htbp]
\renewcommand{\arraystretch}{1.2}
    \centering
        \caption{A snapshot of the dataset. All buggy programs in this example are from project Bugs.jar.}
    \label{tab:dataset}
    \begin{tabular}{l|rrrrcc}
    \toprule
\textbf{Buggy program}&\textbf{wmc}&\textbf{dit}&\textbf{noc}&\textbf{cbo}&\textbf{Kali}&\textbf{Arja}\\\hline
Jackrabbit&9.37&	0.78&	0.23&	12.51&1&0\\
Accumulo &11.94&0.81&0.22&	13.23&1&0\\
Flink&8.43&0.75&	0.31&	10.79&1&1\\
Wicket&8.84&	0.58&	0.41&	11.01&0&1\\
\bottomrule
    \end{tabular}
\end{table}

To create a vector with features for each buggy program, we compute the Object-oriented and Java-Specific method features, which are calculated at the \emph{class-level}. Then we calculate the average value of these features over all classes for each buggy program. 
Next, we compute the \pef{} features. 
Instead of considering all statements from the buggy program, we focus on a subset of them: those that, with a given probability, could have the bug. 
Note that, for predicting which is the most suitable tool given a program bug (Section \ref{sec:mostsuitable}), our approach does not know in which statement(s) the bug is located or the human patch. For this reason we apply fault localisation to filter the statements. To retrieve those statements, we compute the suspicious value of each statement using GZoltar tool~\cite{Campos2012GZoltar}, which uses the Ochiai formula~\cite{Abreu2017Accuracy} to compute the suspiciousness value.
GZoltar is the most prominent fault localisation tool used by the Java repair systems considered in this study. 
For each buggy program, we select the 100 most suspicious statements returned by GZoltar. 
We consider 100 as it is a common cut-off value used in program repair experiment, e.g., see analysis from \cite{Long2016Space}. If a patch we obtain from our dataset is applied in a statement not included in the mentioned list of suspicious statements returned by the fault localisation tool, we include that statement in the list, with the goal of also analysing it. We found 75 bugs having, at least, one patch of such case. Next, we compute the \pef features for each of those statements.  Finally, we compute the average of the features that characterise the suspicious statements.



\section{Results}
\label{sec:results}

We present the results for each research question, and aim to provide insights into why the different APR techniques work. First we present the most significant features that impact APRT effectiveness. Second, we investigate the diversity of exiting buggy datasets used for APR. Next, we investigate the differences between exiting APRTs by analysing their strengths and weaknesses using the most significant features. Finally, we present the results from the Machine Learning algorithms used for APRT selection.

\subsection{RQ1. What impacts the effectiveness of existing APRTs?}

We performed feature learning on the total list of features (described in Section~\ref{sec:method:features}) that were extracted from \textbf{1,282} buggy programs. The aim is to select the best set of features that highlights the strengths and weaknesses of the APR techniques. To account for the randomness in the results, each trial of feature learning was run 10 times on each buggy program for each approach, using different random seeds, and the mean was considered. Out of the 146 features that were part of the study, E-APR identified the following 9 optimal features which best capture the difficulty in generating patches for APR:

(F1) \textbf{MOA:} Measure of Aggregation. 

(F2) \textbf{CAM:} Cohesion Among Methods

(F3) \textbf{AMC:} Average Method Complexity

(F4) \textbf{PMC:} Private Method Count

(F5) \textbf{AECSL:} Atomic Expression Comparison Same Left indicates the number of statements with a binary expression that have more than an atomic expression (e.g., variable access). This feature belongs to \emph{Syntax} category.

(F6) \textbf{SPTWNG:} Similar Primitive Type With Normal Guard indicates the number of statements that contain a variable (local or global) that is also used in another statement contained inside a guard (i.e., an If condition).
This feature belongs to \emph{Usage} category.

(F7) \textbf{CVNI:} Compatible Variable Not Included is the number of local primitive type variables within the scope of a statement that involves primitive variables that are not part of that statement. This feature belongs to \emph{Usage} category.

(F8) \textbf{VCTC:} Variable Compatible Type in Condition measures the number of variables within an If condition that are compatible with another variable in the scope. This feature belongs to \emph{Type} category.

(F9) \textbf{PUIA:} Primitive Used In Assignment measures the number of primitive variables in assignments. 
This feature belongs to \emph{Type} category.

Using these features we were able to define the footprints of the techniques with with the highest topological preservation of 87\% (explained variance). In essence, we can conclude the following.

\vspace{2mm}
\begin{mybox}{mycolor3}

\textbf{RQ1:} The most significant features that have an impact on the effectiveness of APR techniques are the Object-Oriented Features:  MOA, CAM, AMC, PMC, and the \pef feature: AECSL, SPTWNG, CVNI, VCTC, and PUIA.
\end{mybox}

To visualise the results in a 2-D instance space, we apply PCA as a dimensionality reduction technique on the optimal subset of features. Two new axes were created, which are linear combinations of the selected set of most significant features. The coordinate system that defines the new instance space is defined as:

\begin{equation}
\label{eq:pcs}
\begin{bmatrix}
	z_1 \\
	z_2
\end{bmatrix}
= 
 \begin{bmatrix}
	{0.38} & {-0.02}   \\
	 {-0.16} & {0.19} \\
	 {0.37}&{-0.04}\\
	 {-0.06}&{0.36}\\
	 {0.08}&{0.28}\\
	 {0.17}&{0.22}\\
	 {0.07}&{0.31}\\
	 {-0.34}&{0.01}\\
	 {0.12}&{0.16}
\end{bmatrix}^{T}
 \begin{bmatrix}
	{\text{MOA}}  \\
	{\text{AECSL}} \\
	{\text{PMC}}\\
	{\text{SPTWNG}}\\
	{\text{AMC}}\\
	{\text{CVNI}}\\
	{\text{VCTC}}\\
	{\text{CAM}}\\
	{\text{PUIA}}
\end{bmatrix}
\end{equation}

The new coordinates (depicted in Equation~\ref{eq:pcs}) are a combination of the 9 features. CAM, PMC and MOA have the highest contribution on $z_1$, and SPTWNG, AMC and VCTC contribute the most to $z_2$. CVNI, AECSL, and PUIA contribute equally to both coordinates.

\begin{figure}[ht]
    \begin{subfigure}{0.32\linewidth}
    \includegraphics[width=\textwidth]{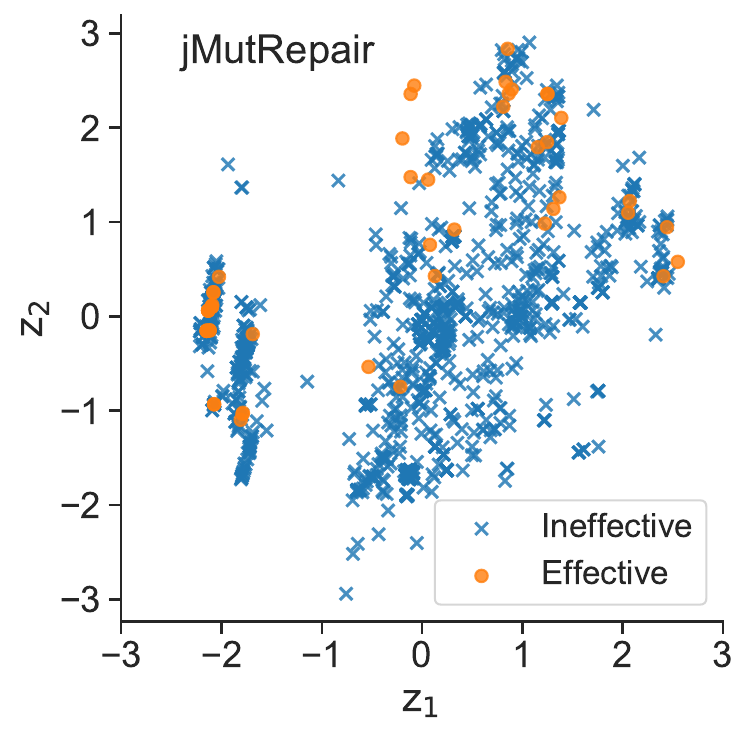}
    \end{subfigure}
    \begin{subfigure}{0.32\linewidth}
    \includegraphics[width=\textwidth]{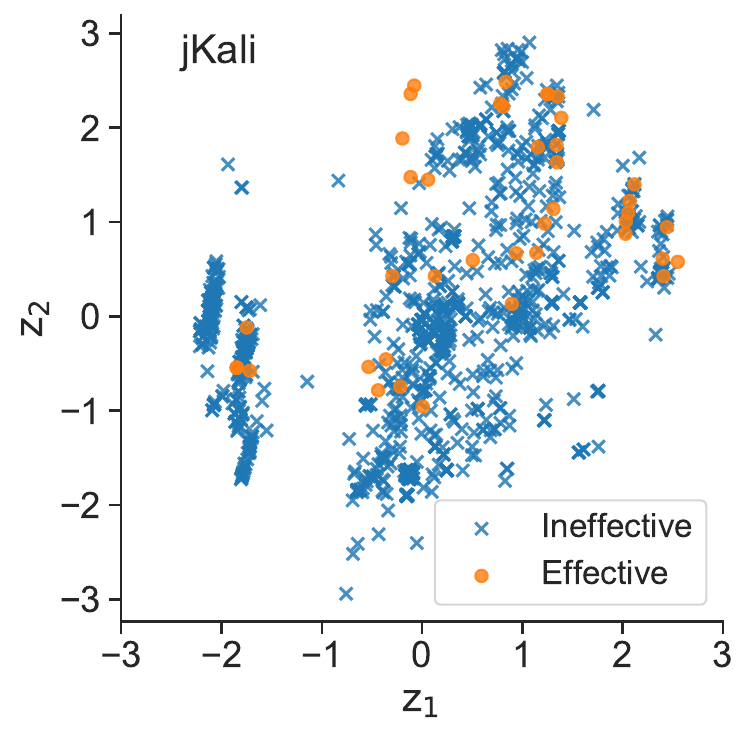}
    \end{subfigure}
    \begin{subfigure}{0.32\linewidth}
    \includegraphics[width=\textwidth]{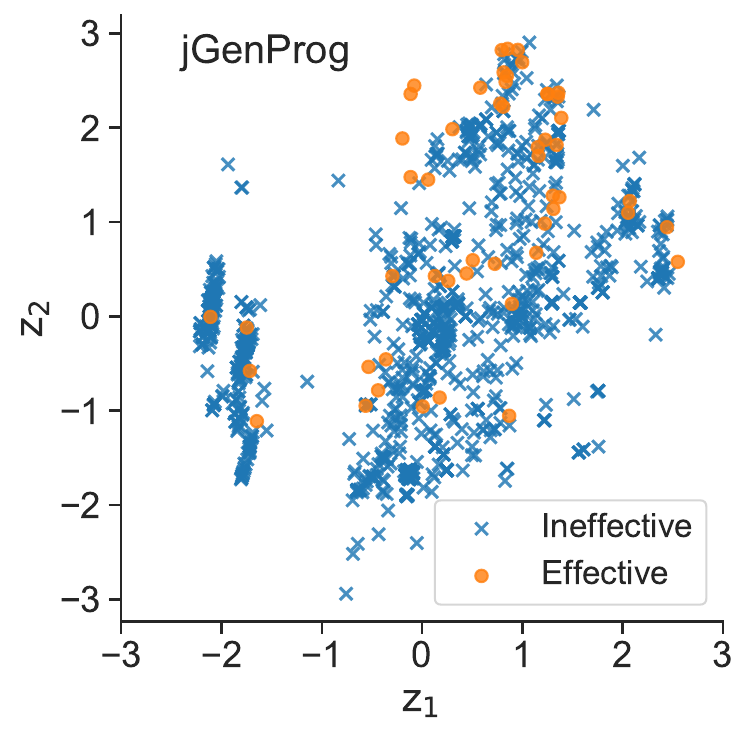}
    \end{subfigure}
    \begin{subfigure}{0.32\linewidth}
    \includegraphics[width=\textwidth]{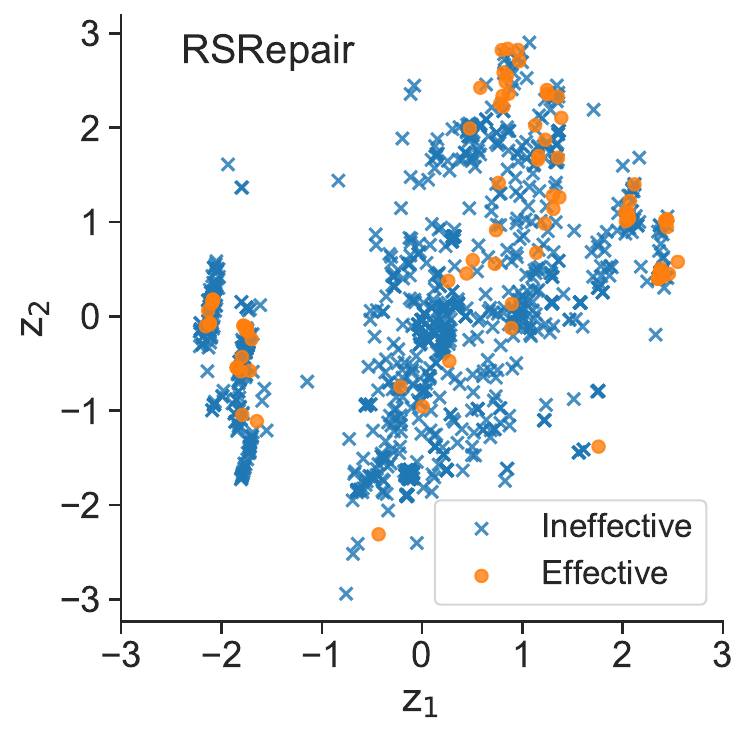}
    \end{subfigure}
    \begin{subfigure}{0.32\linewidth}
    \includegraphics[width=\textwidth]{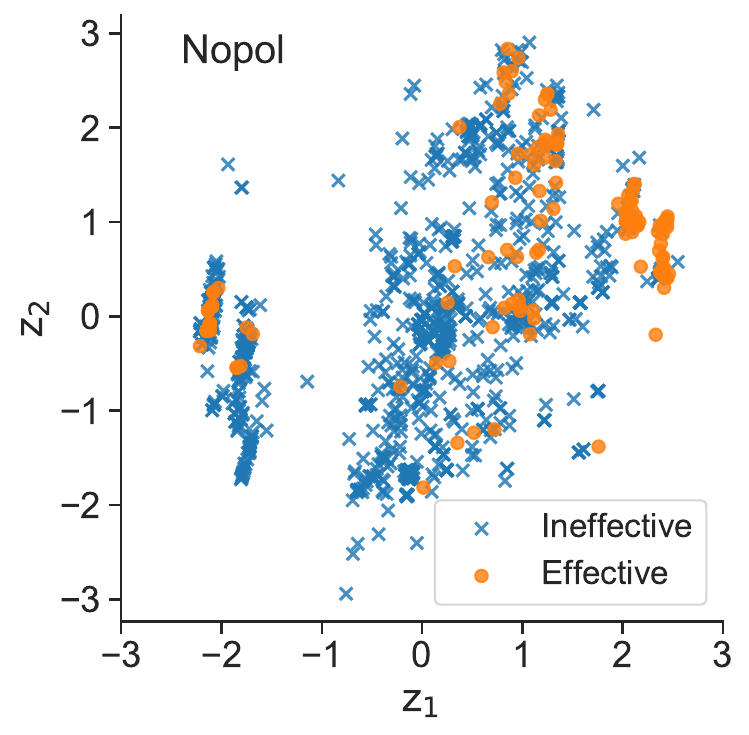}
    \end{subfigure}
    \begin{subfigure}{0.32\linewidth}
    \includegraphics[width=\textwidth]{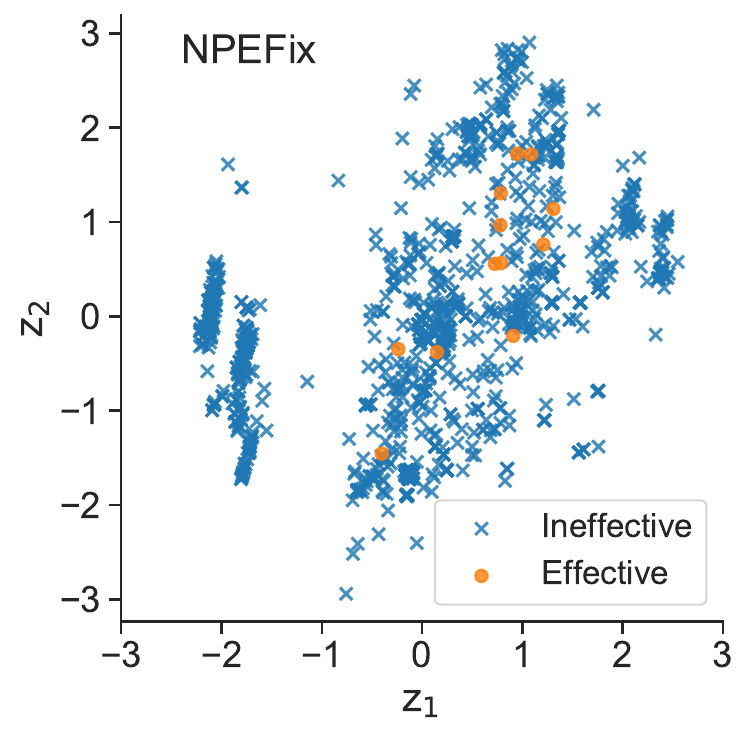}
    \end{subfigure}
    \begin{subfigure}{0.32\linewidth}
    \includegraphics[width=\textwidth]{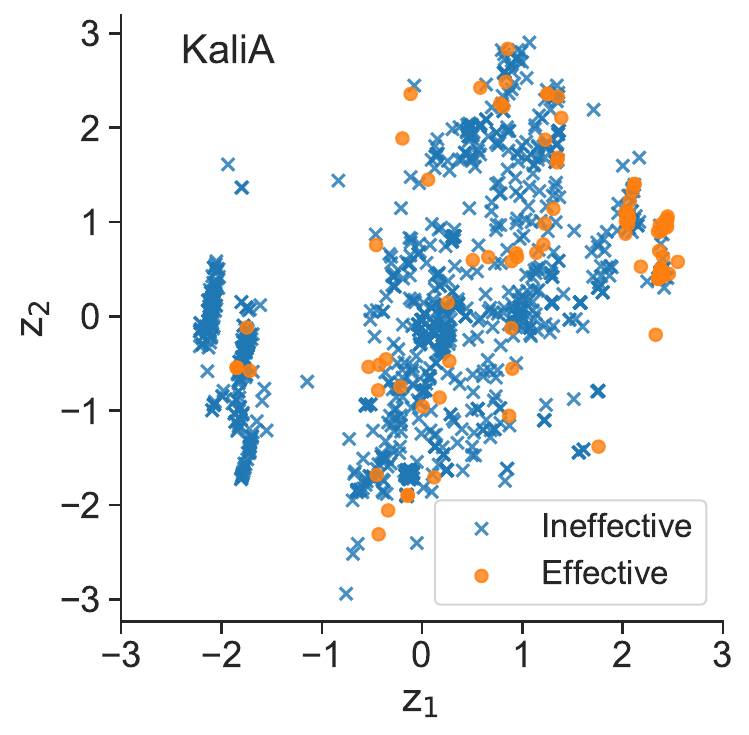}
    \end{subfigure}
        \begin{subfigure}{0.32\linewidth}
    \includegraphics[width=\textwidth]{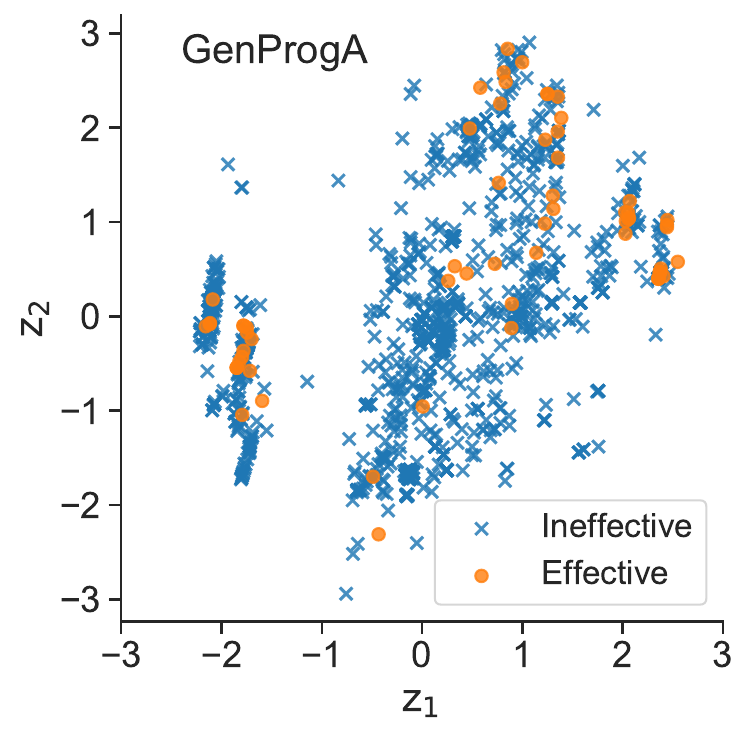}
    \end{subfigure}
    \begin{subfigure}{0.32\linewidth}
    \includegraphics[width=\textwidth]{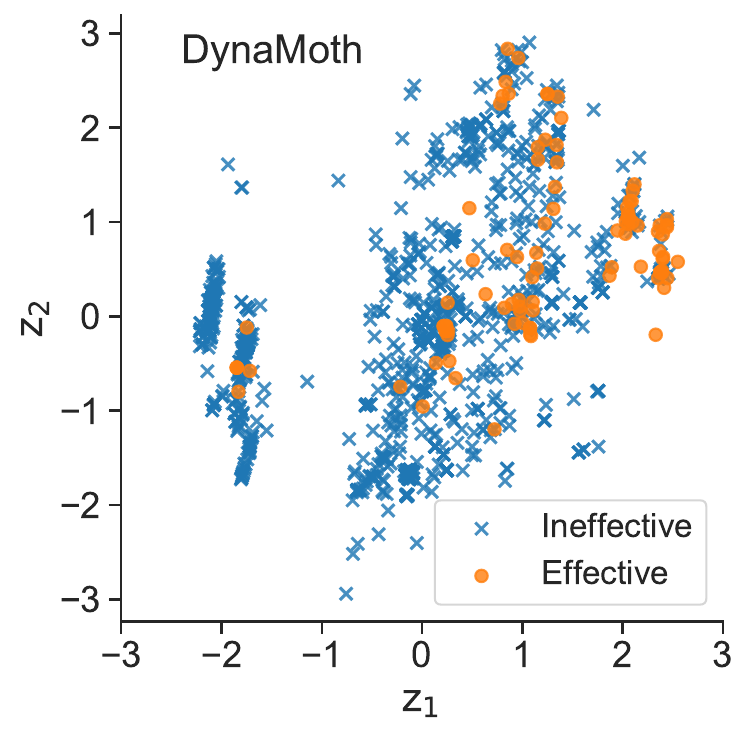}
    \end{subfigure}
    \begin{subfigure}{0.32\linewidth}
    \includegraphics[width=\textwidth]{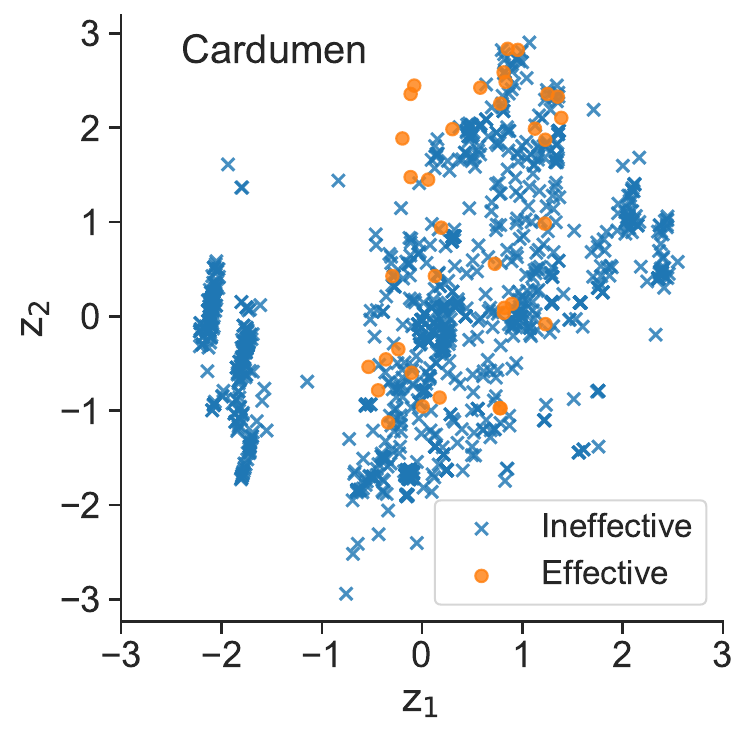}
    \end{subfigure}
    \begin{subfigure}{0.32\linewidth}
    \includegraphics[width=\textwidth]{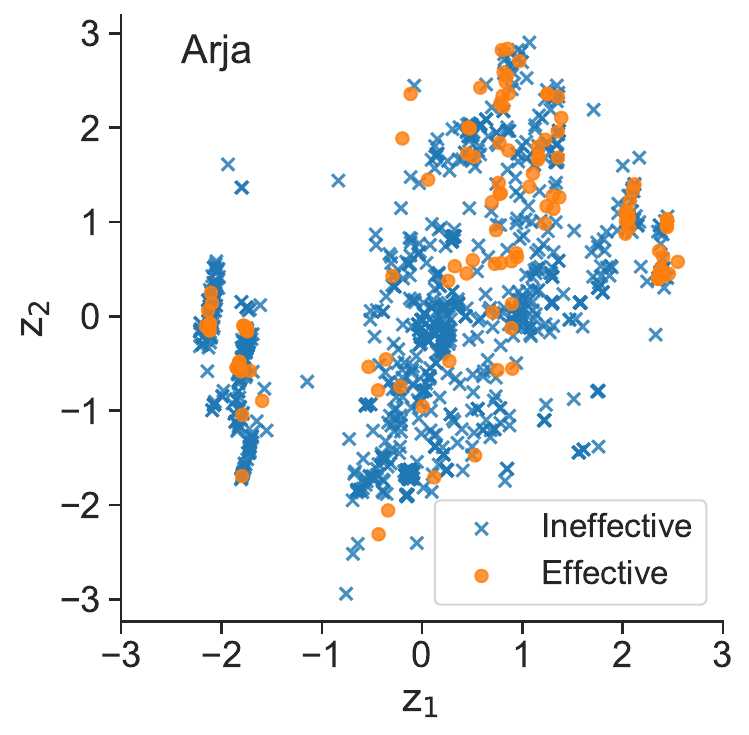}
    \end{subfigure}
    \caption{APR technique footprints. Each point is a buggy class, and is labelled as Effective, if the technique was able to generate a plausible patch for it.}
    \label{fig:footprints}
\end{figure}

We plot the footprints of the 11 APRTs in Figure~\ref{fig:footprints}. Each point in the 2-D instance space represents a buggy program. If an APR technique produced a patch for a particular program, it is considered Effective, otherwise, we label it as Ineffective. Each graph in Figure~\ref{fig:footprints} represents the footprint of one of the techniques that we study in this paper. The x-axis and y-axis are the two principal components $z_1$ and $z_2$, defined in Equation~\ref{eq:pcs}.  

A visual inspection of the footprints shows that while some techniques appear more similar than others (for example, jKali is more similar to jMutRepair than NPEFix), each technique has its unique strengths. 

All APRTs apart form NPEFix repaired bugs located at the top-right of the instance space. These are bugs from Defects4J benchmark (see Figure~\ref{fig:datasetfootprint}), which confirms a long held hypothesis that APRTs are being perfected to repair bugs from this dataset.



\subsubsection{Footprints size}

Table~\ref{tab:relativePrf} shows the area size of the APRT footprints, measured using Equation~\ref{eqFootprintsize}. The size of the footprint is an indication of the overall performance of the APRT. The larger the footprint, the more diverse bugs an APRT can repair.

\begin{table}[!ht]
\renewcommand{\arraystretch}{1.2}
    \centering
        \caption{Performance differences between the APRTs.}
    \label{tab:relativePrf}
    \begin{tabular}{lr|lr}
    \toprule
\begin{tabular}[c]{@{}c@{}} \textbf{APRT}\end{tabular}	&\begin{tabular}[c]{@{}c@{}} \textbf{Footprint size} \end{tabular}&\begin{tabular}[c]{@{}c@{}} \textbf{APRT}\end{tabular}	&\begin{tabular}[c]{@{}c@{}} \textbf{Footprint size} \end{tabular}\\\hline
jMutRepair	& 0.223	&	jKali &	0.215\\
jGenProg	&\textbf{0.388}	&	RSRepair &	0.006 \\
Nopol& 0.236&NPEFix&0\\
KaliA& 0.052&GenProgA&0.004\\
DynaMoth	&0.169		&	Cardumen&0.257	\\
Arja & 0.016\\\bottomrule
    \end{tabular}
\end{table}

While most techniques have relatively similar footprint size, jGenProg is the winner. The footprint size is not based on the number of programs that a technique was able to repair. Instead, the effectiveness of an APRT is measured in terms of the diversity of the features of these programs and their spread in the instance space. An APRT that can repair more diverse bugs is considered to be more effective.

\subsubsection{Significant Software Features}

\begin{figure*}[!ht]
    \begin{subfigure}{0.32\linewidth}
    \includegraphics[width=\textwidth]{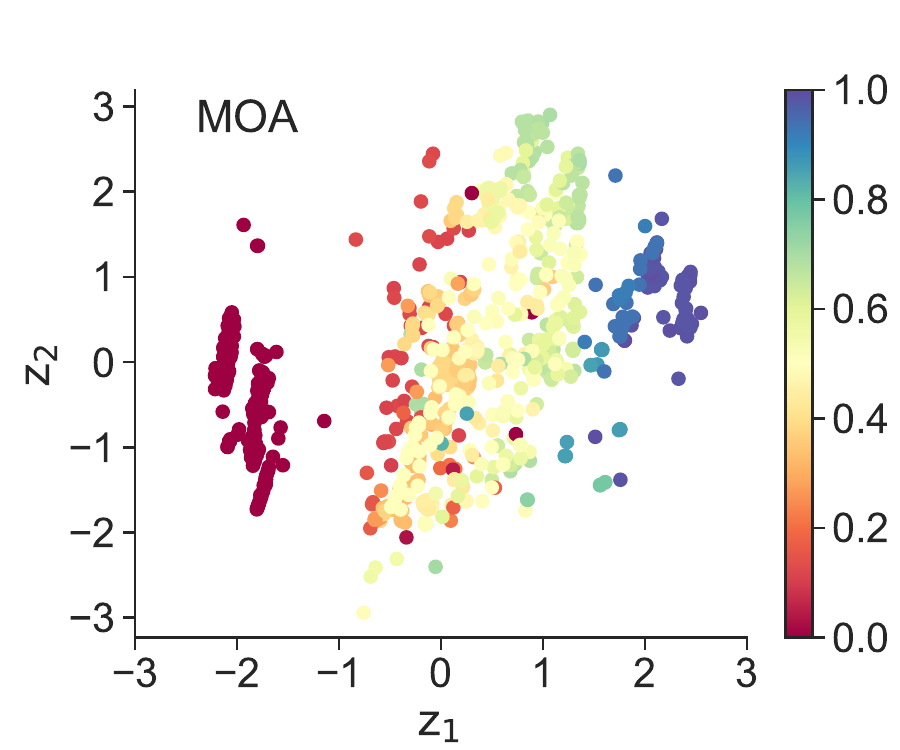}
    \end{subfigure}
    \begin{subfigure}{0.33\linewidth}
    \includegraphics[width=\textwidth]{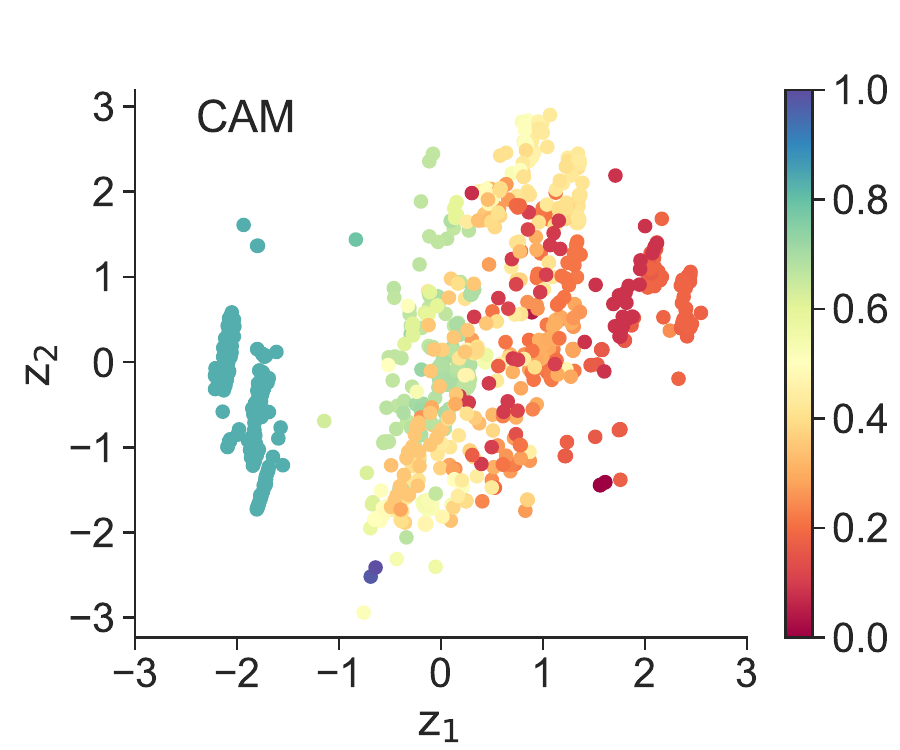}
    \end{subfigure}
        \begin{subfigure}{0.33\linewidth}
    \includegraphics[width=\textwidth]{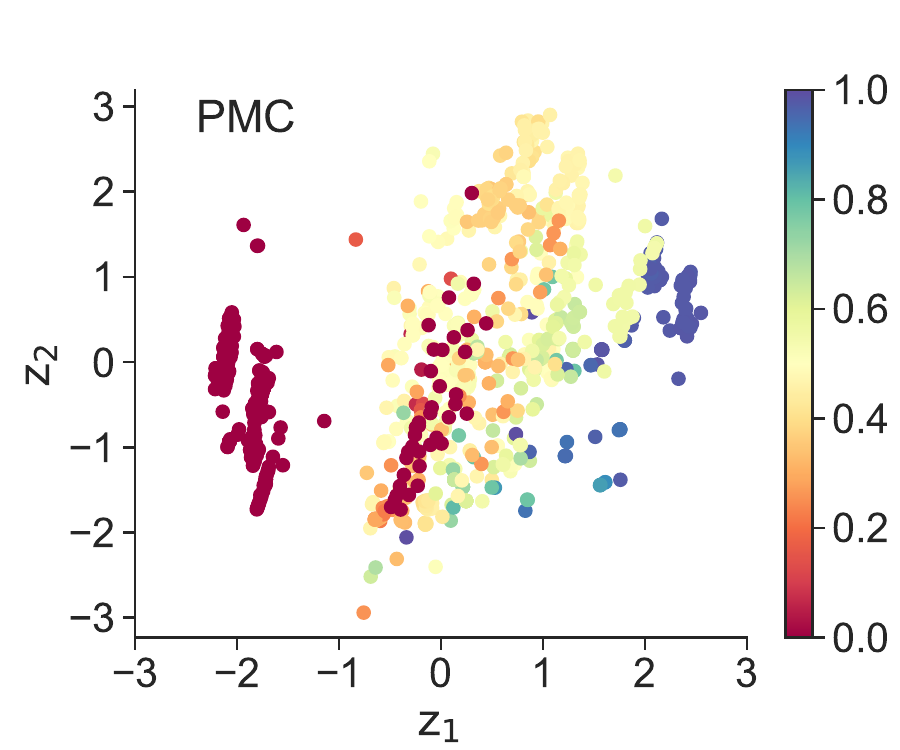}
    \end{subfigure}
            \begin{subfigure}{0.33\linewidth}
    \includegraphics[width=\textwidth]{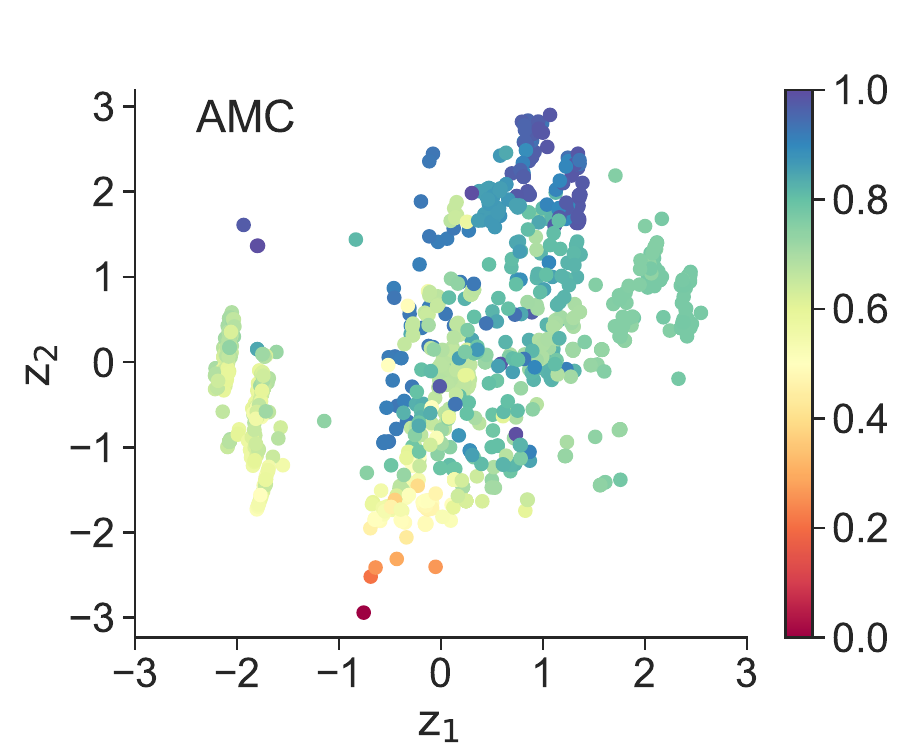}
    \end{subfigure}
    \begin{subfigure}{0.33\linewidth}
    \includegraphics[width=\textwidth]{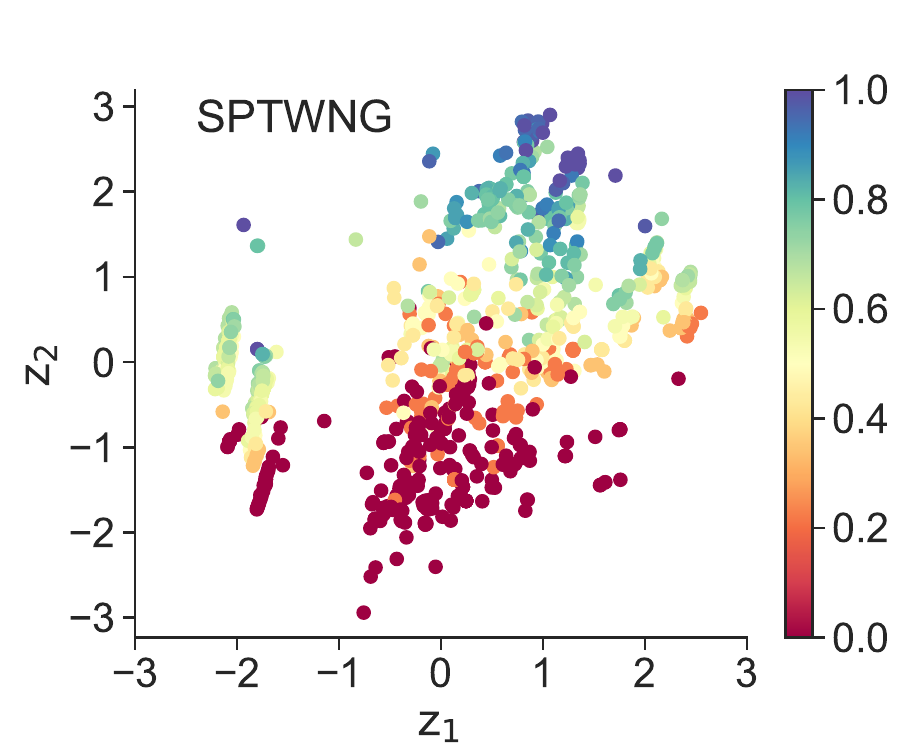}
    \end{subfigure}
            \begin{subfigure}{0.32\linewidth}
    \includegraphics[width=\textwidth]{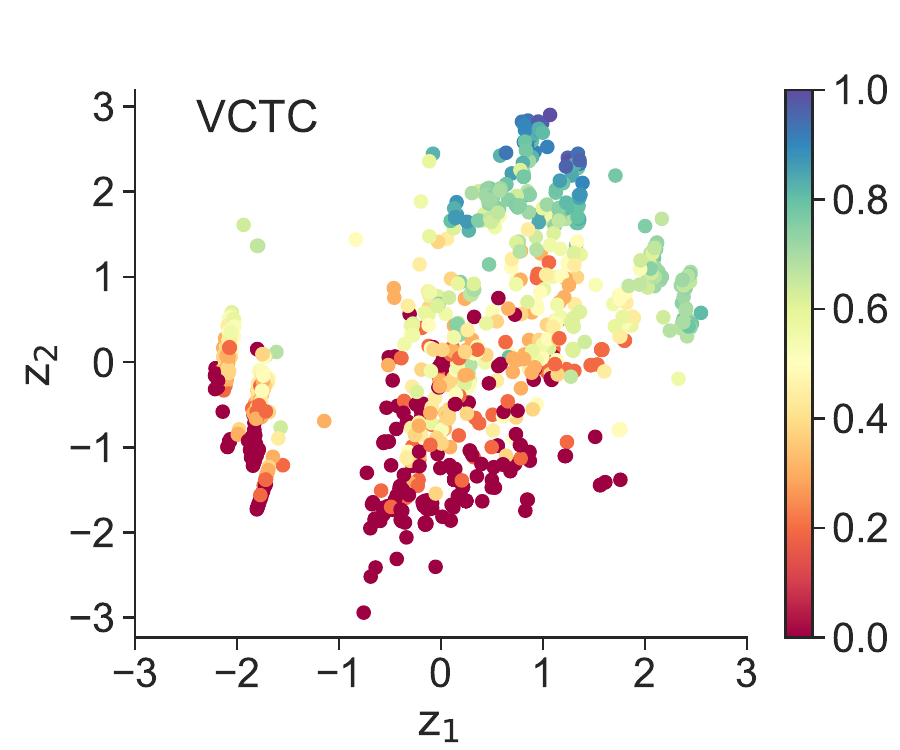}
    \end{subfigure}\\ \vspace{4mm} 
    \begin{subfigure}{0.32\linewidth}
    \includegraphics[width=\textwidth]{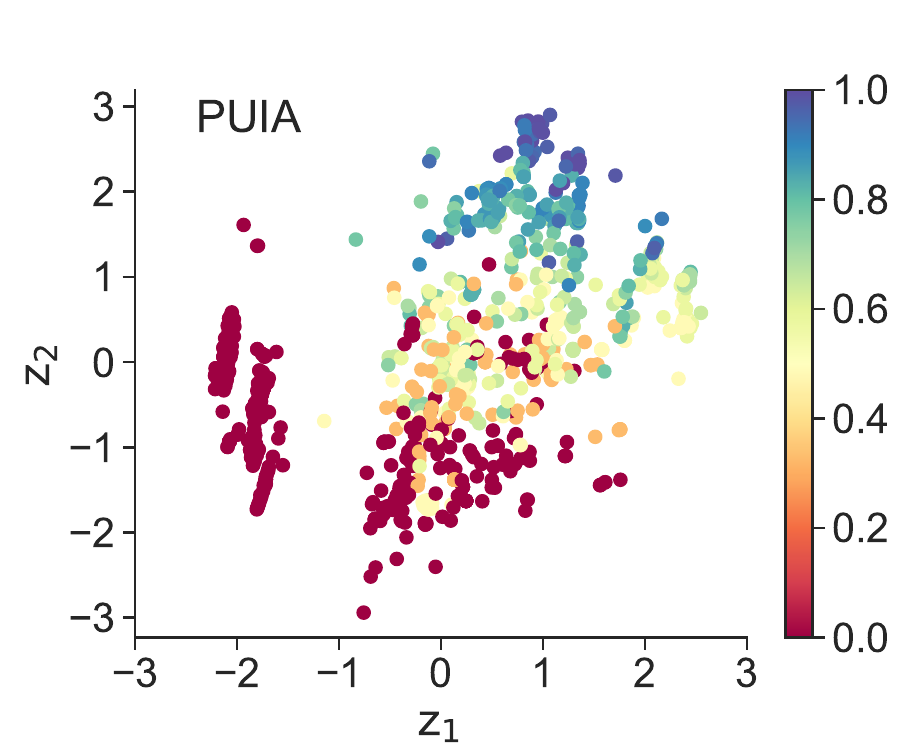}
    \end{subfigure}
    \begin{subfigure}{0.32\linewidth}
    \includegraphics[width=\textwidth]{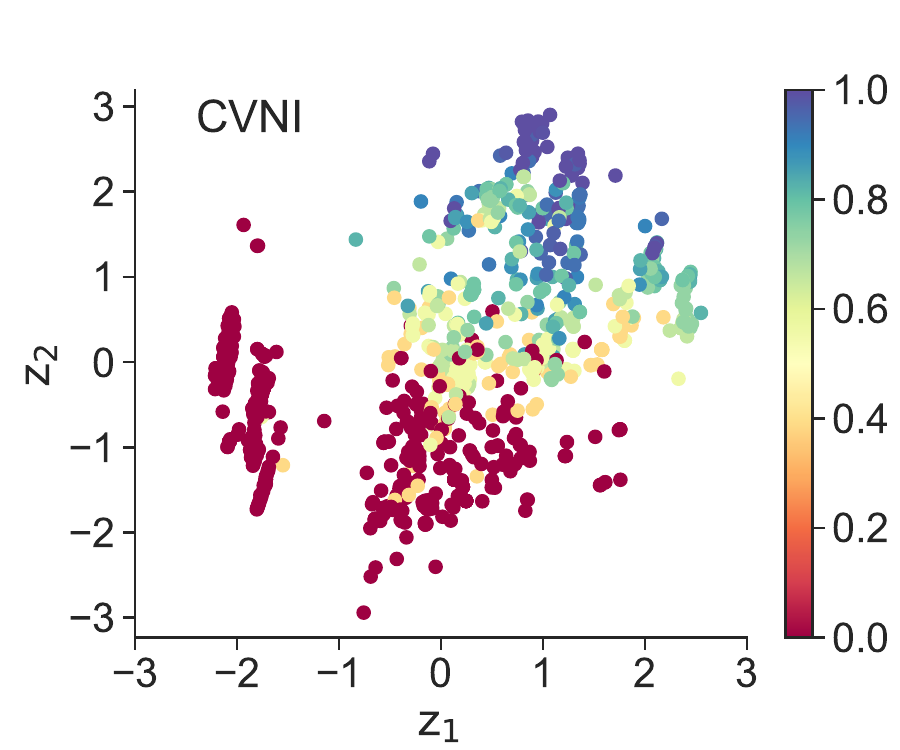}
    \end{subfigure}
        \begin{subfigure}{0.34\linewidth}
    \includegraphics[width=\textwidth]{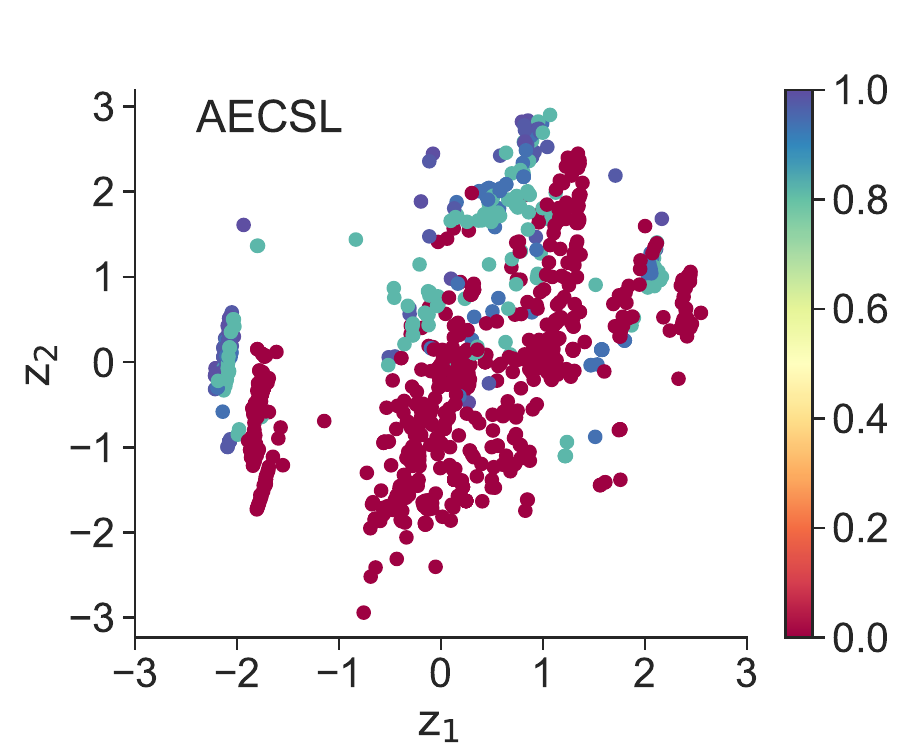}
    \end{subfigure}
    \caption{Feature footprints. The values of features have been normalised between 0 and 1, and the colour scheme is used to represent the values of features.}
    \label{fig:softwareFeatureFootprints}
\end{figure*}

Figure~\ref{fig:softwareFeatureFootprints} depicts the feature footprints, which shows how the buggy program instances score in terms of the most significant features. 

\textbf{MOA and PMC.} The cluster of buggy programs where mostly jMutRepair, RSRepair, Nopol, GenProgA and Arja are effective has a lower measure of aggregation (MOA) and private methods count (PMC). MOA (as defined in Table~\ref{tab:featuresObject}) is the percentage of data declaration in the system whose types are of user defined classes, as opposed to those of system defined classes, such as integers, real numbers etc. It indicates that, compared to other approaches, it is easier for jMutRepair, RSRepair, Nopol, GenProgA and Arja to repair bugs originating from buggy programs that have fewer user declared types and lower number of private methods. jMutRepair, RSRepair, GenProgA and Arja are from the class of generate-and-validate APR techniques, with GenProgA and RSRepair being variations of GenProg. These tools make use of mutation operators to generate patches, which in general can be an effective way to fix bugs, but it proves ineffective in programs with many private methods and user defined types. This indicates that more sophisticated operators are required to fix such programs.

\textbf{CAM.} The third most significant feature is cohesion among methods in a class (CAM), which is a measure of class cohesion. The cluster of buggy programs where mostly jMutRepair, RSRepair, Nopol, GenProgA and Arja are effective is high in terms of CAM. High class cohesion is a desirable property and has previously been linked with high software quality. As mentioned above, MutRepair, RSRepair, GenProgA and Arja use mutation to generate new patches for buggy programs, which is quite simple and works well with highly cohesive programs where related program elements are in the same place (in this case, class). Mutation applies random changes in code, and is less likely to introduce new bugs if classes are highly cohesive. On the other hand, DynaMoth is effective at repairing bugs from programs with low cohesion. DynaMoth is a semantic-based APR tool, which performs a dynamic synthesis of patches for repairing conditional bugs. The tool specifically addresses the issue of complex method calls and low cohesion, which explains its superior performance in buggy programs with low CAM.

\textbf{AMC.} Average Method Complexity is relatively high in the upper right part of the instance space, where most APRTs are able to generate plausible patches. AMC is defined as the average method size (the number of java binary codes in the method) for each class, indicating that APRTs are usually more effective with longer methods.

\textbf{\pef{} features} These metrics capture different characteristics of the buggy parts of the programs. Out of the 146 features, E-APR identified 5 significant \pef{} -- SPTWNG, VCTC, PUIA, CVNI and AECSL -- whose footprints we show in Figure~\ref{fig:softwareFeatureFootprints}. Four of these five features -- SPTWNG, VCTC, PUIA, CVNI -- have very similar footprints.

SPTWNG (Similar Primitive Type With Normal Guard) indicates the number of statements that contain a variable (local or global) that is also used in another statement contained inside a guard (i.e., an If condition). VCTC (Variable Compatible Type in Condition) measures the number of variables within an If condition that are compatible with another variable in the scope. PUIA (Primitive Used In Assignment) measures the number of primitive variables in assignments. CVNI (Compatible Variable Not Included) is the number of local primitive type variables within the scope of a statement that involves primitive variables that are not part of that statement. Finally, AECSL (Atomic Expression Comparison Same Left) indicates the number of statements with a binary expression that have more than an atomic expression (e.g., variable access). Programs with a high value of these features are more likely to be repaired by most techniques, while jMutRepair, Arja, KaliA, Nopol and RSRepair can generate plausible patches even for programs with low feature values. Since these features measure properties of the potential buggy locations, it makes sense that programs with such high feature values are more likely to be repaired.

In summary, the effectiveness of APRTs is impacted by software features, which makes these methods problem dependent, and as such, no technique can be considered the best in all cases. We observe different strengths and weaknesses of existing APRTs, which calls for methods that make it possible to select the most suitable technique given a buggy program with particular features.

\subsection{RQ2. Are existing APR datasets significantly different?}

\begin{figure}[!ht]
    \centering
    \includegraphics[width=0.6\linewidth]{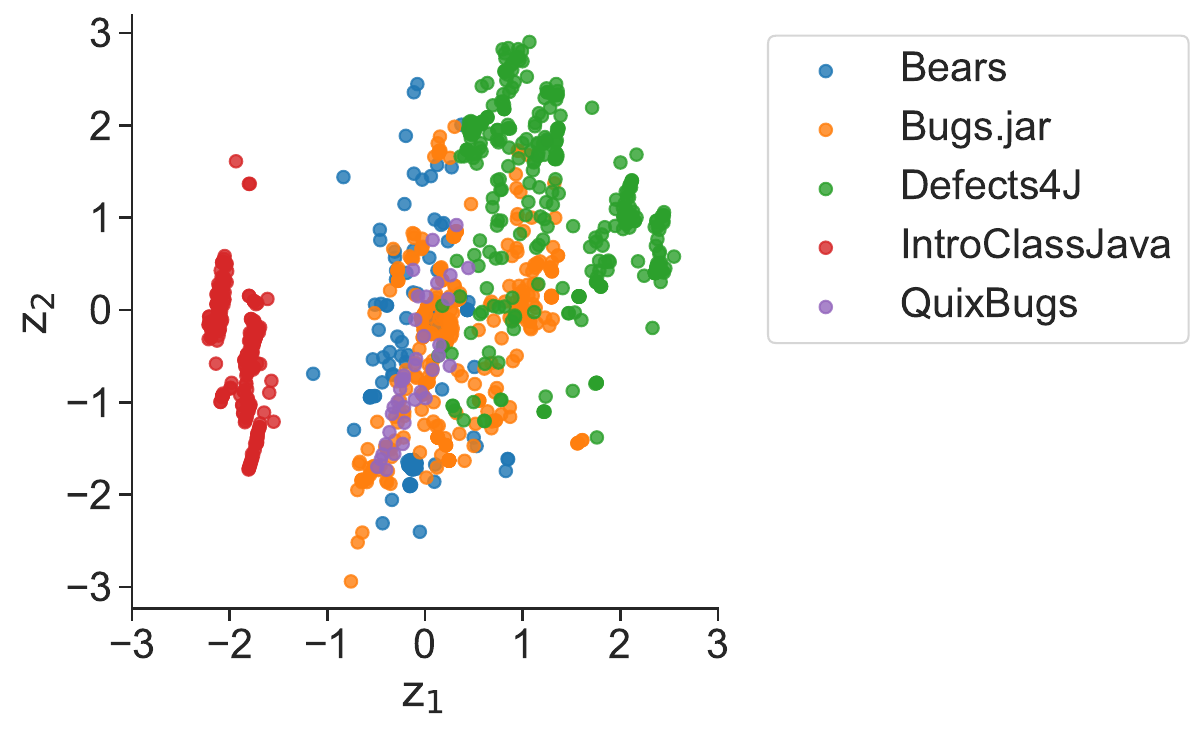}
    \caption{Benchmark footprint. Each point corresponds to a bug from a particular benchmark dataset.}
    \label{fig:datasetfootprint}
\end{figure}

The 2-D instance space that was constructed to analyse the effectiveness of APRTs, also allows us to analyse the location of the different benchmarks, which reveals how diverse they are. The dataset footprint presented in Figure~\ref{fig:datasetfootprint} shows the reduced feature space with instances labelled according to the dataset they belong to. Each point is a bug from a particular dataset.

The features that were eventually found as significant and used to create this instance space, are the ones that have a good linear relationship with algorithm performance. For some APRTs, the choice of features may be better than for others, however, our approach chooses a common feature set that performs well on average across all algorithms.

We observe that there is a distinctive cluster on the left of Figure~\ref{fig:datasetfootprint} composed of only bugs from IntroClassJava. It is clear that this dataset is significantly different from the other datasets. Further away from this cluster, is the footprint of Defects4J, which is on the rightmost side of the graph. This indicates that  Defects4J is significantly different from IntroClassJava. 

On the other hand, the footprints of Bears, Bugs.jar and QuixBugs overlap to a greater extent. 
They are spread between IntroClassJava and Defects4J and have a higher spread than the other datasets. 
Bugs.jar covers a larger are and encompasses that one from Quixbugs.
Bugs.jar contains some bugs obtained from the same software as the other datasets (e.g., Apache, Commons, Math), thus the bugs are eventually the same. 
QuixBugs is a set of buggy implementation of well known algorithms (e.g., Quixsort), and each buggy program in this dataset is a single class. 
The others datasets are real buggy programs, composed of several classes.

In summary, the answer to the second research question is as follows:\\

\begin{mybox}{mycolor3} 

\textbf{RQ2:} IntroClassJava and most of bugs from Defects4J are significantly different from the other benchmark datasets.
Meanwhile, Bugs.jar has the most diverse bugs; its footprint encompasses that of QuixBugs, the majority of Bear's footprint and a portion of Defects4J's footprint.

\end{mybox}

The dataset footprint also helps us understand if a dataset is biased, that is if it doesn't fill the possible instance space, and lies within the `footprint' (area of strength) of one APRT only, and doesn't give other algorithms a chance to show their strength. We particularly observe that the footprint of Defects4J lies within the area of strength of most APRTs apart from NPEFix, whose footprint is shown in Figure~\ref{fig:footprints}. What this means is that 
if the performances of APRTs are compared solely on this dataset, the evaluation can be biased and demonstrate only the strengths of these approaches. The footprints of Bugs.jar, Quixbugs and Bears lie within the area where most APRTs apart from NPEFix are not able to generate plausible patches, indicating that these datasets are quite challenging for these approaches and exhibit less bias.

Our finding from this research task can inform researchers who develop new APRTs in the selection of the bug benchmark to test their technique. It wouldn't be sufficient to test a new APRT on just one dataset, and a technique that works for Defects4J may not produce good results when repairing IntroClassJava.

\subsection{RQ3. How can we select the most suitable APRT?}
\label{sec:mostsuitable}

To answer this question, the E-APR framework uses multi-label classification algorithms to predict the most suitable APRT to repair buggy programs with particular features. We use 10-fold cross validation to evaluate the performance of four notable Machine Learning techniques: Support Vector Machine (SVM), Random Forest Classifier (RFC), Decision Tree (DT), and Multi-Layer Perceptron (MLP). 

We use the scikit-learn Python implementation of these approaches and employed MLSMOTE~\cite{charte2015mlsmote} to address the class imbalance problem. The performance of the two approaches is evaluated in terms of precision, recall, and f1-score. Precision is the fraction of instances that are correctly predicted, calculated as:

\begin{equation}\label{eq:precision}
	P = \frac{TP}{TP+FP}
\end{equation}
where \textit{TP} is the true positives and \textit{FP} is the false positives. Recall measures how accurately the model is able to identify the relevant data.

\begin{equation}\label{eq:rec}
	R = \frac{TP}{TP+FN}
\end{equation}
where FN is false negatives. F1-Score is the harmonic mean of P and R, computed as follows:

\begin{equation}\label{eq:fscore}
	F1 =   2 \frac{P \cdot R}{P + R}
\end{equation}

Results are shown in Table~\ref{tab:MLresults}.

\begin{table}[!ht]
    \centering
    \renewcommand{\arraystretch}{1.2}
    \begin{tabular}{@{}l|rrr|rrr|rrr|rrr@{}}
    \toprule
    &\multicolumn{3}{c|}{\textbf{SVM}}&\multicolumn{3}{c}{\textbf{RFC}}&\multicolumn{3}{c}{\textbf{DT}}&\multicolumn{3}{c}{\textbf{MLP}}\\ \cline{2-13}
    &P &R &F1  &P &R &F1 &P &R &F1&P &R &F1 \\ \hline
Arja & 0.83&  0.76  & 0.80 & 0.89 & \textbf{0.87} & 0.88 &0.96 & 0.84 & 0.90&0.96 & 0.86 & \textbf{0.91} \\
Cardumen & 0.77 & 0.71  & 0.74 &\textbf{0.86}  &  \textbf{0.86}   &  \textbf{0.86} & 0.86   &  0.68 & 0.76 & 0.85 & 0.59& 0.70 \\
DynaMoth & 0.68   &   0.78  &   0.72    & 0.93  & 0.81   & 0.87& 0.83 & 0.81 & 0.82  &\textbf{0.94} &\textbf{0.89}&\textbf{0.91}\\
GenProgA & 0.72   &   \textbf{0.93}  &   0.81    & \textbf{0.86}  &  0.83 & \textbf{0.85} &0.79  &0.81 &0.80& 0.85 &0.59 &0.70\\
KaliA    & \textbf{0.92}   &   0.84  &   0.88    & 0.87  & \textbf{0.91}  & \textbf{0.89}&0.88  & 0.79 & 0.83 &0.88 &0.79 &0.83\\
NPEFix   & 0.47   &   0.82  &   0.60    & \textbf{0.82}  & 0.82 & \textbf{0.82} & 0.56 & \textbf{0.83} &0.67&0.00 & 0.00 &0.00\\
Nopol    & 0.61   &   \textbf{0.93}  &   0.74    & \textbf{0.92} & 0.73  & \textbf{0.81}&0.83 & 0.76 & 0.79 &0.71& 0.45 &0.56\\
RSRepair & \textbf{0.90}   &   0.81  &   \textbf{0.85}    & 0.86 & 0.75  & 0.80& 0.82 &\textbf{0.87} & 0.85 &0.85& 0.74 &0.79\\
jGenProg & 0.65   &   0.42  &   0.51    & \textbf{0.80} & 0.77 & 0.78 & 0.81 & 0.81 & 0.81&0.79 &\textbf{0.83}& 0.81\\
jKali & 0.87 & 0.77  &   0.81    & \textbf{0.93} & 0.88 & \textbf{0.90}&0.86& \textbf{0.91}& 0.89 & 0.89& 0.87& 0.88\\
jMutRepair  & 0.77 &0.48    &   0.59    & \textbf{0.85}  &0.52& 0.65&0.75&0.50& 0.60& 0.84 & \textbf{0.67}& \textbf{0.74}\\\midrule

micro avg  & 0.76 & 0.76   &   0.76    &  \textbf{0.88}& \textbf{0.81}  &\textbf{0.84} & 0.84 & 0.79&    0.82&0.87 &0.74&0.80 \\
macro avg   & 0.74 & 0.75   &   0.73    &  \textbf{0.88}  & \textbf{0.81} &\textbf{0.84}&0.81 &  0.78& 0.79&0.78& 0.67 & 0.72\\
weighted avg& 0.78 & 0.76   &   0.76    &  \textbf{0.88}  &\textbf{0.81} & \textbf{0.84}&0.84& 0.79& 0.81& 0.85& 0.74 &0.79\\\bottomrule

  \end{tabular}
    \caption{The performance of Support Vector Machine (SVM), Random Forest Classifier (RFC), Decision Tree (DT) and Multi-Layer Perceptron (MLP) classifier in terms of precision (P), recall (R) and f1-score (F1).}
    \label{tab:MLresults}
\end{table}

The results indicate that, while all ML algorithms perform well in the task of APRT selection, the performance of RFC is clearly better than SVM. As a comparison, the work from Le et al. \cite{lee2015fixingdelegated} presents a similar task (predict whether a bug can be repaired by a genetic-programming based repair approach i.e., GenProg \cite{LeGoues2012GenProg}) and reports a precision of 72\%.

\vspace{2mm}
\begin{mybox}{mycolor3} 

\textbf{R3:} The Random Forest Classifier is the best performing Machine Learning technique for APRT selection, and can predict the most suitable technique with \textbf{88\%} precision, \textbf{81\%} recall and \textbf{84\%} f1-score. 
\end{mybox}

\vspace{2mm}

Given the high performance of E-APR for predicting the most suitable APRT, it is of high-priority for us to integrate this approach to existing repair infrastructures such as RepairThemAll \cite{Durieux:2019:RepairThemAll} or Repairnator \cite{repairnator2019}. For example, RepairThemAll has 11 automated repair techniques, but it does not offer any capabilities or guidelines in terms of which technique to select. Integrating E-APR with RepairThemAll would make it possible for users to select the most suitable APRT on the fly. 
Repairnator, on the other hand, is a software bot that automatically repairs broken Travis builds. Given a buggy program that produces a build to fail, Repairnator executes different repair approaches (including jGenProg, Nopol, among others) one by one, and the execution order is hard-coded.   By incorporating E-APR, Repairnator could first execute E-APR to obtain the most suitable repair approaches for the buggy program, and execute them accordingly. In the future, we will investigate the effectiveness of integrating our approach with automated program repair infrastructure, such as RepairThemAll and Repairnator. 

The overhead of using E-APR to select the most suitable APRT within existing APR infrastructures is minimal. APRT first must extract the code features by doing static analysis of the buggy program (it takes a few seconds to extract the nine feature we have identified as significant).
Then, based on the extracted features, APRT uses the trained model to perform the prediction in few milliseconds.

\section{Discussion and Threats to Validity}

\subsection{Features selection}

A threat to the validity of this study is the selection of the considered features.
Our approach E-APR considers 3 sets of features, each of them selected with a clear purpose: 
one aims at capturing object-oriented features, the second aims at capturing features specific to Java, and the third aims at capturing features related to the bug fixing activity (\pef features).
Thus, we consider that the set of features is diverse enough to capture the characteristics of the program under analysis.

In this work, we complement extensively used  features (e.g., Object-oriented features \cite{chidamber:1994})  with a novel set of features (\pef features) that aim at characterising buggy programs.
As the latter features are novel, there is a risk they do not precisely characterise buggy programs.
However, a recent work \cite{Yu2019XCRF} used \pef features  to successfully predict source code transformations applied on buggy program.
For this reason, we consider that \pef features can be used to predict the most suitable APR tool to apply to a buggy program.

\subsection{Correctness of patches}

A threat to the validity is the correctness of patches.
In our experiment from Section \ref{sec:results} we consider all generated patches (plausible) rather than focusing only on correct patches. One of the reasons is the availability of data on patch correctness. The number of correct patches generated by APRTs is much lower than the number of plausible patches.
For instance,  the recent manual evaluation done by Trien et al. \cite{tian2020evaluating} of the patches we have used in this paper (more than 67,000 patches from RepairThemAll \cite{Durieux:2019:RepairThemAll})
found only 900 correct patches, from 20 different bugs (14 from Defects4J, 5 from Quixbugs and 1 from Bugs.jar).
We reproduce our experiment, available in our appendix~\cite{appendix}, by considering the bugs that could be repaired by at least one correct patch. We found that the dataset consisting of 20 bugs that were correctly repaired is not sufficient for the machine learning algorithms used to identify significant features and create the algorithm footprints. 

While we think that considering correct patches is an important next step, and a priority for our future work, the results with plausible patches provide some important insights into how APRTs work and how effective they are. 
Current APRTs find it challenging to even produce plausible patches, and E-APR helps understand why this is the case, and what kind of weaknesses future research into APRT should focus on. Moreover, recent studies shows that a plausible patch, even being overfitting and not adequate for repairing a bug, could give developer a valuable piece of information.
For instance, Ginelli et al. \cite{Ginelli2020}  studied code-removal patches, which works on manual machine patch analyses (e.g., \cite{Qi2015Kali}) labelled most of them as overfitting patches. They found that in 95.8\% of the cases having an overfitting code-removal patch, it reveals different kinds of problems affecting the test suites that are relevant for the developers. Thus, they show this type of overfitting patches is useful: it exposes a particular weakness of the test suites.

\subsection{Selection of repair tools}
During the last years, several repair tools have been presented to repair Java bugs.
Two previous works have tried to execute the tools (i.e., the materialization of repair approaches) on real bugs:  \cite{Durieux:2019:RepairThemAll} could executed 11 repairs tools, and \cite{Liu2020EficacyTestSuite}  executed 16 tools.
Both papers list the reasons about why other repair approaches and tools could not be executed.  

In this paper we consider the execution data from 11 tools. The main reason is that those tools were executed on 5 different bug benchmarks and generated patches are publicly available \cite{Durieux:2019:RepairThemAll}. 
Other tools have exclusively focused on Defects4J, and we were not able to generate results for other datasets we consider in this study. 
For example, \cite{Liu2020EficacyTestSuite} evaluated 16 repair approaches only on Defects4J.  
We have included in our appendix \cite{appendix} initial results of an experiment done by considering the patches of that experiment, which includes, in addition to 10 repair tools considered in our paper (all except NPEfix), another 6:  ACS \cite{Xiong2017ACS}, Avatar \cite{liu2019avatar}, FixMiner \cite{Koyuncu2020FixMiner}, kPar \cite{Liu2019FLbias},  SimFix \cite{Jiang2018SimFix}, TBar \cite{Liu2019Tbar}.
From those initial results, we could not draw conclusive results.
Our conjecture is the experiment has not enough diverse data: 
a single dataset evaluated (Defects4J), which contains bugs extracted from only 5 projects (Commons Math, Commons Lang, Joda Time, jFree Chart, and Closure).

We prioritised in this paper having a larger dataset, and found that the techniques we consider are diverse enough to demonstrate the capabilities of the proposed technique.  We consider that this point (i.e., the selection of evaluated tools) does not invalidate the novelty of our technique.

\subsection{Failure information}

Some repair techniques are designed to fix specific bugs, and their effectiveness can be limited by the nature of the bug that is addressed \cite{Monperrus2018bibliography,Gazzola2019survey}. 
For instance, NPEFix repairs null pointer exceptions and is unlikely to be useful in other cases. In this paper, we decided to focus on the features that allow to \emph{characterise} the buggy program under repair, without considering, for instance, the type of failure. Our approach, however, can easily be extended to include failure information. For instance, an extension could include a new set of features that characterise the failure, for example null pointer exceptions, stack overflow, array index error.

E-APR does not consider failure information since most of the bugs considered in this study do not produce a failure, but an \emph{incorrect output}.
The incorrect output is exposed by the failing test case via \emph{assertions}. For instance, by inspecting Defects4J Dissection \cite{Sobreira2018defects4jdissection} we found that 304 out of 395 (77\%) bugs from Defects4J are due to incorrect output. We explored the meta-data of bugs using \url{http://program-repair.org/defects4j-dissection} \cite{Sobreira2018defects4jdissection} and found that 275 bugs are due to an \emph{AssertionFailedError} (for example, Chart-7: \texttt{junit.framework.AssertionFailedError: expected:<1> but was:<3>}) or \emph{unit.framework.ComparisonFailure} (e.g. \texttt{junit.framework.ComparisonFailure: expected:<String[[]]> but was:<String[;]>}). Both exceptions are thrown by the testing framework after detecting the incorrect output.

\subsection{Integration with repair infrastructures}

To our knowledge, repair infrastructures such as Repairnator or RepairThemAll do not have the ability to predict, given a buggy program taken as input, with is the most suitable APR tool to generate a test-suite adequate patch for it.
Repairnator calls APR tools in a fixed order, independently of the characteristics of the program under analysis.
On the contrary, the user of RepairThemAll must decide the APR to be call.
In both cases, our approach E-APR could be integrated to both of them.
For Repairnator, E-APR could determine the order of the APR tools to be called with the goal of calling first the tools that are most suitable for a given buggy program.
Similarly, for RepairThemAll,  E-APR could suggest the user the APR tool to be invoked.

\section{Related Work on the Effectiveness of APR Techniques}

Researchers working in the area of APR have acknowledged that evaluating the quality of patches produced by APR techniques is crucial~\cite{Martinez2017experiment,Smith2015}. To this end, Qi et al. \cite{Qi2015Kali} studied the plausible generated by GenProg \cite{LeGoues2012Study} for C programs, and classified them as plausible (passing all tests),  \textit{overfitting} (plausible and incorrect) and  \textit{correct} (plausible, and do not have latent defects and do not introduce new defects or vulnerabilities \cite{Long2016Space}). They found that most of the reported patches were \textit{overfitting}. 
Long and Rinard  \cite{Long2016Space} analysed the patch search space of two repair tools, SPR \cite{Long2015SPR} and Prophet \cite{Long2016prophet}, and found that overfitting patches are typically orders of magnitude more abundant than correct patches.

Other works have studied the ability of APR techniques to repair buggy Java programs. For example, Martinez et al. \cite{Martinez2017experiment} manually studied the correctness of patches produced by three APR techniquess over defects from Defects4J benchmark. They found that only a small number of bugs (9/47) could be correctly repaired.
%
Liu et al. \cite{Liu2020EficacyTestSuite} executed 16 repairs tools on Defects4J and manually analyzed the generated patches following the procedure defined in that work.
 They found that the percentage of patches correctness varies between the tools at is belong of the 37\% for 15/16 tools.
Ye et al. \cite{Ye2019StudyQuixBugs} studied the repairability of bugs from QuixBugs \cite{lin2017QuixBugs}, a dataset of 40 small buggy programs (between 9 and 69 LOC).
They found that 15 bugs could be repaired by Nopol \cite{Xuan16MDCLDLM} and approaches from Astor \cite{Martinez2016Astor}, which generated in total 64 plausible patches. However, they found that 33 of them were incorrect. 
%


The presence of overfitting patches has motivated researchers to investigate the amount of the overfitting patches (e.g., \cite{Yu2019Alleviating,Wang2020AutomatedPC}), detect overfitting patches (e.g., DiffTGen \cite{issta17-difftgen}, PatchSim \cite{Xiong2018Identifiying}, Static code feature via learning \cite{Wang2020AutomatedPC}, ODS~\cite{Ye2019ODS}), and  to avoid generating such patches (e.g., UnsatGuided \cite{Yu2019Alleviating}, CapGen \cite{Wen2018CapGen}, Anti-pattern \cite{anti-pattern}).
Empirical studies also have studied overfitting patches in detail.
For example, Liu et al \cite{Liu2020EficacyTestSuite} conducted an large-scale empirical study which analyzed the correctness of patches generated by 16 repairs tools (10 mentioned in Section \ref{sec:all_approaches} (all except NPEFix) and ACS \cite{Xiong2017ACS}, Avatar \cite{liu2019avatar}, FixMiner \cite{Koyuncu2020FixMiner}, kPar \cite{Liu2019FLbias},  SimFix \cite{Jiang2018SimFix}, TBar \cite{Liu2019Tbar}). 
One of their main findings is that many plausible patches are related to wrong locations of the patches. 
As previously found by Liu et al \cite{Liu2019FLbias}, the accuracy of fault localization tool has a direct and substantial impact on the performance of APR tools.

Our work extends existing research in analysing the effectiveness of APR techniques by examining what software features impacts the repairability of a software system. We characterise a software system using code features (e.g., depth of inheritance tree and method cohesion) and determine the most significant features that have impact on whether an APR technique can generate a patch.

There has also been some research in characterising patches generated by APR techniques to investigate how these patches differ from the ones generated by human programmers. 
Wang et al. \cite{1906.03447} compared the difference between 177 correct patches for Defects4J bugs generated by APR techniques and the patches written by developers. To characterise the bugs, the authors considered 6 metrics: a) Patch size, b) Number of chunks c) Number of modified files, d) Number of modified methods e) Line coverage, and f) Branch coverage. They found that automatically generated patches are on average syntactically different compared to the patches generated by developers. Patches generated by APR techniques are usually longer, have a higher number of chunks, and have a higher line and branch coverage.

Similarly, Smith et al. \cite{Smith2015} studied the quality of patches generated by two C program repair approaches (GenProg and TprAutoRepair). 
The authors used two metrics that were \emph{dynamically} computed (i.e., by running the program under repair): a) number of passing and failing test cases, and b) test suite coverage.

Both Wang et al. \cite{1906.03447} and Smith et al. \cite{Smith2015} focus on analysing the kind of patches generated by APR techniques. The aim of these works is to understand how good the patches are, and how they are different from developer-generated patches. Our work, instead, aims at understanding what kind of software systems and bugs APR techniques are able to repair. This will help explain how and why they work, and as a result, make it possible to select the right technique given a new buggy software system. 

In their research, Smith et al. \cite{Smith2015} state that ``Automatic repair should be used in the appropriate contexts'' and that ``Our results suggest that more work is needed to fully understand and characterise test suite quality beyond coverage metrics alone''. The E-APR framework addresses these two research challenges by investigating 146 features, and building a machine learning model that enables the selection of the most suitable APR technique for a given buggy program.

Another related work is the one by Motwani et al.~\cite{Motwani2018evaluation} which investigates correlations between the effectiveness of APR techniques and different aspects of bugs, such as bug importance and bug complexity. Results were analysed at course-grained level, with the findings showing weak to moderate correlation between bug importance and the ability of the APR technique to produce a patch. The results also show that APR techniques are effective in repairing easy bugs - as measured by the number of files and lines that have to be changed to fix the bug - while struggling with more complex bugs. This study makes an important step towards understanding where APR techniques work. In this paper, we take this research one step further by providing a more detailed analysis of the effectiveness of different APR techniques. The framework we propose allows us to examine the effectiveness of individual techniques in a visual and numerical way. We measure the footprints of the different APR techniques and whether their results overlap. This helps us understand the strengths and weaknesses of individual techniques, and their similarities in a more fine-grained way.    

Le et al. \cite{lee2015fixingdelegated} present a work that has a similar goal to ours: they build an oracle that can predict whether fixing a failure should be delegated to a genetic-programming-based automated repair technique. The authors first extract features from an early stage of running a repair tool. 
Then, they pass the values of these features to learn a discriminative model capable of predicting whether continuing a genetic programming search will lead to a repair within a desired time limit.
Beyond the similarities, there are notable differences between our work and the work by Le et al. \cite{lee2015fixingdelegated}. 

First, Le et al. \cite{lee2015fixingdelegated} focus on genetic-programming-based  automated repair technique, while our approach is independent of the type of repair technique. For instance, it considers genetic-programming-based technique (Arja and Genprog), semantic-based techniques (Nopol) and exhaustive methods (jMutRepair). Le et al. \cite{lee2015fixingdelegated} consider 27 features, 18 of which are related to genetic-programming. We use 3 sets of features (in total more than 200 features) that are independent of any repair technique and aim to describe the buggy program under repair. Le et al. \cite{lee2015fixingdelegated} analyse the early stage of a genetic-programming-based technique to extract 18 features. This means that it could be necessary to modify a repair approach to extract those features or to monitor the execution logs.
E-APR considers the features extracted from a buggy program and trains the prediction model using the output from previous executions (i.e., a bug was patched or not by a technique). Le et al. \cite{lee2015fixingdelegated} considers one dataset of bugs (ManyBugs \cite{LeGoues2015ManyBugsIntroClass} with 105 bugs) and one repair tool (GenProg), while we consider 11 repair tools and 5 datasets (1282 bugs).

Lin et al \cite{Lin2020Understanding} studied the non-repairability factors of various APR techniques. They analysed 11,818 execution logs from 27 Java tools, and found that 25.7\% of them contained unexpected exceptions that prevent those tools to find a patch.

\section{Conclusion}
In this paper, we introduced E-APR, which is a novel framework for assessing strengths and weaknesses of APR techniques for Automated Program Repair (APR). We identified nine significant software features that have an impact on APRT effectiveness. These features were then used to provide explanations on an APR technique's effectiveness across a range of buggy programs. We introduced a method for visualising APRT footprints, which reveal strengths and weaknesses of the APR techniques in fixing buggy programs. 

We conducted an analysis of 11 different APR techniques applied to 2,141 bugs from 130 projects, constituting in total 23,551 repair attempts. Our approach effectively identified APRT footprints and the features that impact the effectiveness of an automated program technique. Using the most significant features, we applied two machine learning approaches that learns the relationship between software features and APRT effectiveness. Random Forest Classifier showed the best performance, with 88\% precision, 81\% recall and 84\% f1-score. 

 \section*{Acknowledgement}

The authors would like to acknowledge Prof. Kate Smith-Miles and her team working on  \url{matilda.unimelb.edu.au}. The methodology on Instance Space Analysis constitutes the foundations of this work. Matilda was used to create the instance spaces presented in Figures~\ref{fig:footprints} and~\ref{fig:softwareFeatureFootprints}.

\bibliographystyle{IEEEtran}
\bibliography{references}

\end{document}